\documentclass[12pt]{article}

\usepackage{amsmath}
\usepackage{amssymb}

\allowdisplaybreaks

\begin{document}

\baselineskip=18.8pt plus 0.2pt minus 0.1pt

\makeatletter

\@addtoreset{equation}{section}
\renewcommand{\theequation}{\thesection.\arabic{equation}}
\renewcommand{\thefootnote}{\fnsymbol{footnote}}
\newcommand{\beq}{\begin{equation}}
\newcommand{\eeq}{\end{equation}}
\newcommand{\bea}{\begin{eqnarray}}
\newcommand{\eea}{\end{eqnarray}}
\newcommand{\nn}{\nonumber\\}
\newcommand{\hs}[1]{\hspace{#1}}
\newcommand{\vs}[1]{\vspace{#1}}
\newcommand{\Half}{\frac{1}{2}}
\newcommand{\p}{\partial}
\newcommand{\ol}{\overline}
\newcommand{\wt}[1]{\widetilde{#1}}
\newcommand{\ap}{\alpha'}
\newcommand{\bra}[1]{\left\langle  #1 \right\vert }
\newcommand{\ket}[1]{\left\vert #1 \right\rangle }
\newcommand{\vev}[1]{\left\langle  #1 \right\rangle }
\newcommand{\ul}[1]{\underline{#1}}

\makeatother

\begin{titlepage}
\title{
\vspace{1cm}
On First Order Symmetry Operators for the Field Equations of Differential Forms
}
\author{Yoji Michishita
\thanks{
{\tt michishita@edu.kagoshima-u.ac.jp}
}
\\[7pt]
{\it Department of Physics, Faculty of Education, Kagoshima University}\\
{\it Kagoshima, 890-0065, Japan}
}

\date{\normalsize August, 2020}
\maketitle
\thispagestyle{empty}

\begin{abstract}
\normalsize
We consider first order symmetry operators for the equations of motion of differential $p$-form fields
in general $D$-dimensional background geometry of any signature for both massless and massive cases. 
For $p=1$ and $p=2$ we give the general forms of the symmetry operators.
Then we find a class of symmetry operators for arbitrary $p$ and $D$, which is naturally suggested by
the lower $p$ results.
\end{abstract}
\end{titlepage}

\clearpage
\section{Introduction}

In field theories we have to deal with various types of fields such as scalars, spinors, vectors, etc. 
Especially in supergravity theories we have the metric, Rarita-Schwinger fields, and differential forms.
We often need to solve the linearized equations of motion for those fields.
If we write the equation of motion for a field $\phi$ schematically in the form of $M\phi=0$,
$M$ is a derivative operator of at most second order.
When we solve this equation in given background geometries, it is useful to consider symmetry operators i.e.
a pair of operators $(\mathcal{Q},\mathcal{S})$ satisfying $\mathcal{Q}M=M\mathcal{S}$.
$\mathcal{S}$ generates new solution $\mathcal{S}\phi$ from a solution $\phi$,
and if $\mathcal{Q}=\mathcal{S}$, $\mathcal{S}$ commutes with $M$ and
we can take simultaneous eigenfunctions of $M$ and $\mathcal{S}$.

Among such operators first order ones are especially simple and useful. Therefore we consider the case where 
$\mathcal{Q}$ and $\mathcal{S}$ are first order. (If $\mathcal{S}$ is first order and $\mathcal{Q}$ is higher,
it is also useful, but we do not consider such possibility in this paper.) First order symmetry operators 
for spinors, Rarita-Schwinger fields and the metric perturbation have been considered in
\cite{ms79,bc96,bk04,ckk11,m18,m19}. 

In this paper we consider first order symmetry operators for differential $p$-form fields 
both for massless and massive cases. 
For recent related works for $p=1$ case see 
\cite{ab16,a16,a17,l17,fkk18,kfk18,fkks18,fk18,hty19,hty19-2}, and for arbitrary $p$ see \cite{l19}.
We do not impose the background equation of motion on the background geometry and the metric can have any signature.
In section 2 we give the conditions for first order symmetry operators, and
in section 3 we show how we can extract information on the symmetry operators from those conditions.
In section 4 we present the solutions of the full conditions for $p=1$ and $p=2$, and a partial solution for $p=3$.
Those solutions suggest a class of solutions for arbitrary $p$, and
in section 5 we show it indeed satisfies all the conditions.
Most of our symmetry operators consist of (conformal) Killing-Yano forms, and trivial parts such as
gauge transformation for massless case.
In Appendix we briefly summarize 
properties of conformal Killing-Yano forms used in the calculations.

\section{Preliminaries}

We consider a $p$-form field $A_{\mu_1\dots\mu_p}$ in a $D$-dimensional background space with the metric $g_{\mu\nu}$.
$p$ is in the interval $1\le p\le D-1$. No assumption about the signature of $g_{\mu\nu}$ is made. 

The standard Lagrangian for $A_{\mu_1\dots\mu_p}$ (up to the overall constant factor) is given by 
\bea
\mathcal{L} & = & \sqrt{|g|}\Big(\frac{p+1}{2p!}A_{\mu_1\dots\mu_p}\nabla_\lambda\nabla^{[\lambda}A^{\mu_1\dots\mu_p]}
 -\frac{1}{2p!}m^2A_{\mu_1\dots\mu_p}A^{\mu_1\dots\mu_p}\Big)
\nn & = &
 \frac{1}{p!}\sqrt{|g|}A^{\mu_1\dots\mu_p}M_{\mu_1\dots\mu_p}{}^{\nu_1\dots\nu_p}A_{\nu_1\dots\nu_p},
\eea
where $m$ is the mass parameter and 
\bea
M_{\mu_1\dots\mu_p}{}^{\nu_1\dots\nu_p} & = & 
 \delta_{[\mu_1}{}^{\nu_1}\dots\delta_{\mu_p]}{}^{\nu_p} (\nabla^2-m^2)
 -p\delta_{[\mu_1}{}^{[\nu_1}\dots\delta_{\mu_{p-1}}{}^{\nu_{p-1}}\nabla^{\nu_p]}\nabla_{\mu_p]} 
\label{leqop}
\eea
is a hermitian operator. We do not impose any background field equation on $g_{\mu\nu}$.

The equation of motion for $A_{\mu_1\dots\mu_p}$ is 
$M_{\mu_1\dots\mu_p}{}^{\nu_1\dots\nu_p}A_{\nu_1\dots\nu_p}=0$.
We consider symmetry operators for $M_{\mu_1\dots\mu_p}{}^{\nu_1\dots\nu_p}$ i.e. 
operators $\mathcal{Q}_{\mu_1\dots\mu_p}{}^{\nu_1\dots\nu_p}$ and 
$\mathcal{S}_{\mu_1\dots\mu_p}{}^{\nu_1\dots\nu_p}$ satisfying
\beq
\mathcal{Q}_{\mu_1\dots\mu_p}{}^{\rho_1\dots\rho_p}M_{\rho_1\dots\rho_p}{}^{\nu_1\dots\nu_p}
=M_{\mu_1\dots\mu_p}{}^{\rho_1\dots\rho_p}\mathcal{S}_{\rho_1\dots\rho_p}{}^{\nu_1\dots\nu_p}.
\label{symopcond0}
\eeq
This kind of operators help solving the equation of motion.
In this paper we deal with first order operators i.e. 
$\mathcal{Q}_{\mu_1\dots\mu_p}{}^{\nu_1\dots\nu_p}$ and $\mathcal{S}_{\mu_1\dots\mu_p}{}^{\nu_1\dots\nu_p}$
in the following form:
\bea
\mathcal{Q}_{\mu_1\dots\mu_p}{}^{\nu_1\dots\nu_p} & = &
 Q_{\mu_1\dots\mu_p}{}^{\nu_1\dots\nu_p\lambda}\nabla_\lambda + q_{\mu_1\dots\mu_p}{}^{\nu_1\dots\nu_p},
\\ 
\mathcal{S}_{\mu_1\dots\mu_p}{}^{\nu_1\dots\nu_p} & = &
 S_{\mu_1\dots\mu_p}{}^{\nu_1\dots\nu_p\lambda}\nabla_\lambda + s_{\mu_1\dots\mu_p}{}^{\nu_1\dots\nu_p}.
\eea

We have some examples of the symmetry operators which can be easily found.
The first example is the unit matrix:
\beq
\mathcal{S}_{\mu_1\dots\mu_p}{}^{\nu_1\dots\nu_p} = \mathcal{Q}_{\mu_1\dots\mu_p}{}^{\nu_1\dots\nu_p}
 = c \delta_{[\mu_1}{}^{\nu_1}\dots\delta_{\mu_p]}{}^{\nu_p},
\quad c= \text{const.},
\label{trivial0}
\eeq
and the second example is given by the gauge transformation for the massless case $m=0$:
\bea
\mathcal{S}_{\mu_1\dots\mu_p}{}^{\nu_1\dots\nu_p} & = &
 U_{[\mu_2\dots\mu_p}{}^{\nu_1\dots\nu_p}\nabla_{\mu_1]}
 + \nabla_{[\mu_1}U_{\mu_2\dots\mu_p]}{}^{\nu_1\dots\nu_p}, 
\\
\mathcal{Q}_{\mu_1\dots\mu_p}{}^{\nu_1\dots\nu_p} & = & 
 V_{\mu_1\dots\mu_p}{}^{[\nu_2\dots\nu_p}\nabla^{\nu_1]},
\eea
where $U_{\mu_1\dots\mu_{p-1}}{}^{\nu_1\dots\nu_p}$ and $V_{\mu_1\dots\mu_p}{}^{\nu_1\dots\nu_{p-1}}$ are
arbitrary tensors (except that the first $p-1$ indices and the rest of $U_{\mu_1\dots\mu_{p-1}}{}^{\nu_1\dots\nu_p}$ are
antisymmetrized, and the first $p$ indices and the rest of $V_{\mu_1\dots\mu_p}{}^{\nu_1\dots\nu_{p-1}}$ are
antisymmetrized.)
These give eigenvectors of zero eigenvalue for $M_{\mu_1\dots\mu_p}{}^{\nu_1\dots\nu_p}$.
The third example is given by the isometry of the background with a Killing vector $K^\mu$:
\bea
\mathcal{S}_{\mu_1\dots\mu_p}{}^{\nu_1\dots\nu_p}A_{\nu_1\dots\nu_p}
 & = & \mathcal{Q}_{\mu_1\dots\mu_p}{}^{\nu_1\dots\nu_p}A_{\nu_1\dots\nu_p}
\nn & = & \mathcal{L}_KA_{\mu_1\dots\mu_p}
\nn & = & 
 \Big(\delta_{[\mu_1}{}^{\nu_1}\dots\delta_{\mu_p]}{}^{\nu_p}K^\lambda\nabla_\lambda
 + p\delta_{[\mu_1}{}^{[\nu_1}\dots\delta_{\mu_{p-1}}{}^{\nu_{p-1}}\nabla_{\mu_p]}K^{\nu_p]} \Big)
 A_{\nu_1\dots\nu_p},
\eea
where $\mathcal{L}_K$ is the Lie derivative operator with respect to $K^\mu$.
We will see these examples appear as part of our symmetry operators later.

Let us write down the condition \eqref{symopcond0} for each order of derivative.
Note that indices on covariant derivatives can be symmetrized by the followings:
\bea
\nabla^{\lambda_1}\nabla^{\lambda_2} A_{\mu_1\dots\mu_p}
 & = & \nabla^{(\lambda_1}\nabla^{\lambda_2)} A_{\mu_1\dots\mu_p}
 +\frac{p}{2}R^{\lambda_1\lambda_2}{}_{[\mu_1}{}^\rho A_{|\rho|\mu_2\dots\mu_p]},
\\
\nabla^{\lambda_1}\nabla^{\lambda_2}\nabla^{\lambda_3} A_{\mu_1\dots\mu_p}
& = & \nabla^{(\lambda_1}\nabla^{\lambda_2}\nabla^{\lambda_3)} A_{\mu_1\dots\mu_p}
 + \frac{2}{3}R^{\lambda_1(\lambda_2\lambda_3)\rho}\nabla_\rho A_{\mu_1\dots\mu_p}
\nn & &
 + pR^{\lambda_1(\lambda_2}{}_{[\mu_1}{}^{|\rho|}\nabla^{\lambda_3)} A_{|\rho|\mu_2\dots\mu_p]}
 + \frac{p}{2}R^{\lambda_2\lambda_3}{}_{[\mu_1}{}^\rho\nabla^{\lambda_1} A_{|\rho|\mu_2\dots\mu_p]}
\nn & &
 + \frac{p}{2}\nabla^{\lambda_1}R^{\lambda_2\lambda_3}{}_{[\mu_1}{}^\rho A_{|\rho|\mu_2\dots\mu_p]}
 - \frac{p}{3}\nabla^{(\lambda_2}R^{\lambda_3)\lambda_1}{}_{[\mu_1}{}^\rho A_{|\rho|\mu_2\dots\mu_p]}.
\eea
Then terms proportional to $\nabla_{(\lambda_1}\nabla_{\lambda_2}\nabla_{\lambda_3)}A_{\nu_1\dots\nu_p}$, 
$\nabla_{(\lambda_1}\nabla_{\lambda_2)}A_{\nu_1\dots\nu_p}$, $\nabla_{\lambda}A_{\nu_1\dots\nu_p}$, and 
$A_{\nu_1\dots\nu_p}$ in \eqref{symopcond0}
cancel separately.
From the part proportional to $\nabla_{(\lambda_1}\nabla_{\lambda_2}\nabla_{\lambda_3)}A_{\nu_1\dots\nu_p}$,
\bea
\lefteqn{ Q_{\mu_1\dots\mu_p}{}^{\nu_1\dots\nu_p(\lambda_1}g^{\lambda_2\lambda_3)}
 - pQ_{\mu_1\dots\mu_p}{}^{[\nu_1\dots\nu_{p-1}|(\lambda_1\lambda_2} g^{\lambda_3)|\nu_p]}
} \nn & = &
 S_{\mu_1\dots\mu_p}{}^{\nu_1\dots\nu_p(\lambda_1}g^{\lambda_2\lambda_3)}
 - pS_{[\mu_1\dots\mu_{p-1}}{}^{(\lambda_1|\nu_1\dots\nu_p|\lambda_2}\delta_{\mu_p]}{}^{\lambda_3)}.
\label{Qo3}
\eea
From the part proportional to $\nabla_{(\lambda_1}\nabla_{\lambda_2)}A_{\nu_1\dots\nu_p}$,
\bea
\lefteqn{ q_{\mu_1\dots\mu_p}{}^{\nu_1\dots\nu_p}g^{\lambda_1\lambda_2}
 - pq_{\mu_1\dots\mu_p}{}^{[\nu_1\dots\nu_{p-1}|(\lambda_1} g^{\lambda_2)|\nu_p]}
} \nn & = & 
 s_{\mu_1\dots\mu_p}{}^{\nu_1\dots\nu_p}g^{\lambda_1\lambda_2}
 - ps_{[\mu_1\dots\mu_{p-1}}{}^{(\lambda_1|\nu_1\dots\nu_p|}\delta_{\mu_p]}{}^{\lambda_2)}
\nn & &
 +2\nabla^{(\lambda_1}S_{\mu_1\dots\mu_p}{}^{|\nu_1\dots\nu_p|\lambda_2)}
 -p\nabla_\rho S^\rho{}_{[\mu_2\dots\mu_p}{}^{\nu_1\dots\nu_p(\lambda_1}\delta_{\mu_1]}{}^{\lambda_2)}
 -p\nabla_{[\mu_1}S^{(\lambda_1}{}_{\mu_2\dots\mu_p]}{}^{|\nu_1\dots\nu_p|\lambda_2)}.
\label{qo2}
\eea
From the part proportional to $\nabla_\lambda A_{\nu_1\dots\nu_p}$,
\bea
\lefteqn{ -\frac{2}{3}R^{\rho\lambda} Q_{\mu_1\dots\mu_p}{}^{\nu_1\dots\nu_p}{}_\rho
 -pR^\lambda{}_{\rho\sigma}{}^{[\nu_1} Q_{\mu_1\dots\mu_p}{}^{|\sigma|\nu_2\dots\nu_p]\rho}
 -\frac{1}{3}pR_\rho{}^{[\nu_1}{}_\sigma{}^{|\lambda} Q_{\mu_1\dots\mu_p}{}^{\sigma|\nu_2\dots\nu_p]\rho}
} \nn \lefteqn{
 -\frac{1}{6}pR_{\rho\sigma}{}^{\lambda[\nu_1} Q_{\mu_1\dots\mu_p}{}^{|\sigma|\nu_2\dots\nu_p]\rho}
 +\frac{1}{2}pR_\rho{}^{[\nu_1} Q_{\mu_1\dots\mu_p}{}^{|\lambda|\nu_2\dots\nu_p]\rho}
 -\frac{1}{2}pR_\rho{}^{[\nu_1} Q_{\mu_1\dots\mu_p}{}^{|\rho|\nu_2\dots\nu_p]\lambda}
} \nn \lefteqn{
 -\frac{1}{4}p(p-1)R_{\rho\sigma}{}^{[\nu_1\nu_2} Q_{\mu_1\dots\mu_p}{}^{|\lambda\sigma|\nu_3\dots\nu_p]\rho}
 +\frac{1}{4}p(p-1)R_{\rho\sigma}{}^{[\nu_1\nu_2} Q_{\mu_1\dots\mu_p}{}^{|\rho\sigma|\nu_3\dots\nu_p]\lambda}
} \nn \lefteqn{
 -\frac{1}{4}p(p-1)R_{\sigma\tau\rho}{}^{[\nu_1} Q_{\mu_1\dots\mu_p}{}^{|\sigma|\nu_2\dots\nu_{p-1}|\tau\rho|}
 g^{\nu_p]\lambda}
 -m^2 Q_{\mu_1\dots\mu_p}{}^{\nu_1\dots\nu_p\lambda}
} \nn & = & 
 (\nabla^2-m^2)S_{\mu_1\dots\mu_p}{}^{\nu_1\dots\nu_p\lambda}
 -p\nabla^\rho\nabla_{[\mu_1} S_{|\rho|\mu_2\dots\mu_p]}{}^{\nu_1\dots\nu_p\lambda}
\nn & & 
 +\frac{1}{3}R^\lambda{}_\rho S_{\mu_1\dots\mu_p}{}^{\nu_1\dots\nu_p\rho}
 +pR^{\lambda\rho}{}_\sigma{}^{[\nu_1} S_{\mu_1\dots\mu_p}{}^{|\sigma|\nu_2\dots\nu_p]}{}_\rho
\nn & & 
 -\frac{1}{3}pR^\rho{}_{[\mu_1|\sigma}{}^\lambda S_{\rho|\mu_2\dots\mu_p]}{}^{\nu_1\dots\nu_p\sigma}
 -\frac{1}{3}pR^\rho{}_{\sigma[\mu_1}{}^\lambda S_{|\rho|\mu_2\dots\mu_p]}{}^{\nu_1\dots\nu_p\sigma}
\nn & & 
 -\frac{1}{2}p^2R^\rho{}_{[\mu_1|\sigma}{}^{[\nu_1} S_{\rho|\mu_2\dots\mu_p]}{}^{|\sigma|\nu_2\dots\nu_p]\lambda}
 -\frac{1}{2}p^2R_{[\mu_1|\rho\sigma|}{}^{[\nu_1} S^{|\lambda}{}_{\mu_2\dots\mu_p]}{}^{\sigma|\nu_2\dots\nu_p]\rho}
\nn & & 
 -\frac{1}{2}p^2R^\rho{}_{\sigma\tau}{}^{[\nu_1} S_{\rho[\mu_2\dots\mu_p}{}^{|\tau|\nu_2\dots\nu_p]\sigma}
  \delta_{\mu_1]}{}^\lambda
\nn & & 
 +2\nabla^\lambda s_{\mu_1\dots\mu_p}{}^{\nu_1\dots\nu_p}
 -p\nabla^\rho s_{\rho[\mu_2\dots\mu_p}{}^{\nu_1\dots\nu_p}\delta_{\mu_1]}{}^\lambda
 -p\nabla_{[\mu_1}s^\lambda{}_{\mu_2\dots\mu_p]}{}^{\nu_1\dots\nu_p}.
\label{O1}
\eea
From the part proportional to $A_{\nu_1\dots\nu_p}$,
\bea
\lefteqn{ \frac{1}{3}p\nabla^{[\nu_1}R_{\rho\sigma}Q_{\mu_1\dots\mu_p}{}^{|\rho|\nu_2\dots\nu_p]\sigma}
 - \frac{2}{3}p\nabla_\sigma R_\rho{}^{[\nu_1}Q_{\mu_1\dots\mu_p}{}^{|\rho|\nu_2\dots\nu_p]\sigma}
} \nn \lefteqn{ 
 + \frac{1}{4}p(p-1)\nabla_\sigma R_{\rho\tau}{}^{[\nu_1\nu_2}Q_{\mu_1\dots\mu_p}{}^{|\rho\tau|\nu_3\dots\nu_p]\sigma}
} \nn \lefteqn{ 
 - \frac{1}{2}pR_\rho{}^{[\nu_1}q_{\mu_1\dots\mu_p}{}^{|\rho|\nu_2\dots\nu_p]}
 + \frac{1}{4}p(p-1)R_{\rho\sigma}{}^{[\nu_1\nu_2}q_{\mu_1\dots\mu_p}{}^{|\rho\sigma|\nu_3\dots\nu_p]}
 - m^2q_{\mu_1\dots\mu_p}{}^{\nu_1\dots\nu_p}
} \nn & = & 
 \frac{2}{3}p\nabla_\rho R^{\sigma[\nu_1}S_{\mu_1\dots\mu_p}{}^{|\rho|\nu_2\dots\nu_p]}{}_\sigma
 - \frac{2}{3}p\nabla^{[\nu_1}R_\rho{}^{|\sigma|}S_{\mu_1\dots\mu_p}{}^{|\rho|\nu_2\dots\nu_p]}{}_\sigma
\nn & & 
 - \frac{2}{3}p^2\nabla^\rho R_{[\mu_1|\sigma\tau}{}^{[\nu_1}S_{\rho|\mu_2\dots\mu_p]}{}^{|\tau|\nu_2\dots\nu_p]\sigma}
 - \frac{1}{3}p^2\nabla_\sigma R^\rho{}_{[\mu_1|\tau}{}^{[\nu_1}S_{\rho|\mu_2\dots\mu_p]}{}^{|\tau|\nu_2\dots\nu_p]\sigma}
\nn & & 
 + p\nabla^\rho S_{\mu_1\dots\mu_p}{}^{\tau[\nu_2\dots\nu_p|\sigma|}R_{\rho\sigma\tau}{}^{\nu_1]}
 - \frac{1}{2}p^2\nabla^\rho S_{\rho[\mu_2\dots\mu_p}{}^{\tau[\nu_2\dots\nu_p|\sigma|}R_{\mu_1]\sigma\tau}{}^{\nu_1]}
\nn & & 
 - \frac{1}{2}p^2\nabla_{[\mu_1}S^\rho{}_{\mu_2\dots\mu_p]}{}^{\tau[\nu_2\dots\nu_p|\sigma|}R_{\rho\sigma\tau}{}^{\nu_1]}
\nn & & 
 + (\nabla^2-m^2)s_{\mu_1\dots\mu_p}{}^{\nu_1\dots\nu_p}
 - p\nabla^\rho\nabla_{[\mu_1}s_{|\rho|\mu_2\dots\mu_p]}{}^{\nu_1\dots\nu_p}
\nn & & 
 - \frac{1}{2}p^2s_{\rho[\mu_2\dots\mu_p}{}^{\tau[\nu_2\dots\nu_p}R^{|\rho|}{}_{\mu_1]\tau}{}^{\nu_1]}.
\label{O0}
\eea
We have to solve these four conditions. \eqref{Qo3} is purely algebraic and does not contain derivatives.
It restricts the tensor structure of $Q_{\mu_1\dots\mu_p}{}^{\nu_1\dots\nu_p\lambda}$ and
$S_{\mu_1\dots\mu_p}{}^{\nu_1\dots\nu_p\lambda}$. 
Then \eqref{qo2} restricts the structure of $q_{\mu_1\dots\mu_p}{}^{\nu_1\dots\nu_p}$ and
$s_{\mu_1\dots\mu_p}{}^{\nu_1\dots\nu_p}$. \eqref{O1} and \eqref{O0} give further restrictions.

\section{Analysis of the conditions for the symmetry operators}

In this section we show the details of the procedure for obtaining information on the solution to 
the conditions \eqref{Qo3} and \eqref{qo2}. 
We show how $Q_{\mu_1\dots\mu_p}{}^{\nu_1\dots\nu_p\lambda}$,  
$q_{\mu_1\dots\mu_p}{}^{\nu_1\dots\nu_p}$ and
$s_{\mu_1\dots\mu_p}{}^{\nu_1\dots\nu_p}$ are determined from $S_{\mu_1\dots\mu_p}{}^{\nu_1\dots\nu_p\lambda}$.
For $S_{\mu_1\dots\mu_p}{}^{\nu_1\dots\nu_p\lambda}$,
since full analysis for arbitrary $p$ is complicated, we only show how to determine part of the full structure.
The rest of the structure can be determined in a similar way. 
The results in this section will be necessary in the following sections.
Since analysis of the conditions \eqref{O1} and \eqref{O0} for arbitrary $p$ is similar to the one we will show
in section 5, we do not show its details here.

\paragraph{Analysis of \eqref{Qo3}}
First let us consider \eqref{Qo3}.
By contracting $\lambda_2$ and $\lambda_3$ in \eqref{Qo3}, we obtain
\bea
\lefteqn{
(D-p+1)Q_{\mu_1\dots\mu_p}{}^{\nu_1\dots\nu_p\lambda}
+(p+1)Q_{\mu_1\dots\mu_p}{}^{[\nu_1\dots\nu_p\lambda]}
-pQ_{\mu_1\dots\mu_p}{}^{\rho[\nu_2\dots\nu_p}{}_\rho g^{\nu_1]\lambda}}
\nn & = & 
 (D-p+2)S_{\mu_1\dots\mu_p}{}^{\nu_1\dots\nu_p\lambda}
-pS^\lambda{}_{[\mu_2\dots\mu_p}{}^{\nu_1\dots\nu_p}{}_{\mu_1]}
-pS_{\rho[\mu_2\dots\mu_p}{}^{\nu_1\dots\nu_p\rho}{}\delta_{\mu_1]}{}^\lambda.
\label{Qo3-1}
\eea
By antisymmetrizing indices $\nu_1,\dots,\nu_p$, and $\lambda$ in this equation,
\bea
\lefteqn{Q_{\mu_1\dots\mu_p}{}^{[\nu_1\dots\nu_p\lambda]} }
\nn & = &
 \frac{D-p+2}{D+2}S_{\mu_1\dots\mu_p}{}^{[\nu_1\dots\nu_p\lambda]}
-\frac{p}{D+2}S^{[\lambda}{}_{[\mu_2\dots\mu_p}{}^{\nu_1\dots\nu_p]}{}_{\mu_1]}
-\frac{p}{D+2}S_{\rho[\mu_2\dots\mu_p}{}^{[\nu_1\dots\nu_p|\rho|}{}\delta_{\mu_1]}{}^{\lambda]}.
\label{Qo3-1-1}
\eea
Then from the above two equations,
\bea
Q_{\mu_1\dots\mu_p}{}^{\nu_1\dots\nu_p\lambda} & = &
 \frac{p}{D-p+1}Q_{\mu_1\dots\mu_p}{}^{\rho[\nu_2\dots\nu_p}{}_\rho g^{\nu_1]\lambda}
\nn & & 
 + \frac{D+1}{(D+2)(D-p+1)}\Big[
  (D-p+2)S_{\mu_1\dots\mu_p}{}^{\nu_1\dots\nu_p\lambda}
\nn & & 
  -pS^\lambda{}_{[\mu_2\dots\mu_p}{}^{\nu_1\dots\nu_p}{}_{\mu_1]}
  -pS_{\rho[\mu_2\dots\mu_p}{}^{\nu_1\dots\nu_p\rho}{}\delta_{\mu_1]}{}^\lambda
 \Big]
\nn & & 
 + \frac{p}{(D+2)(D-p+1)}\Big[
  (D-p+2)S_{\mu_1\dots\mu_p}{}^{[\nu_1\dots\nu_{p-1}|\lambda|\nu_p]}
\nn & & 
  -pS^{[\nu_1}{}_{[\mu_2\dots\mu_p}{}^{|\lambda|\nu_2\dots\nu_p]}{}_{\mu_1]}
  -pS_{\rho[\mu_2\dots\mu_p}{}^{[\nu_1\dots\nu_{p-1}|\lambda\rho|}{}\delta_{\mu_1]}{}^{\nu_p]}
 \Big].
\label{Qo3-2}
\eea
By applying this to the terms on the left hand side of \eqref{Qo3}, 
we obtain an equation consisting of only $S_{\mu_1\dots\mu_p}{}^{\nu_1\dots\nu_p\lambda}$.
Therefore \eqref{Qo3-2} contains all the information on $Q_{\mu_1\dots\mu_p}{}^{\nu_1\dots\nu_p\lambda}$.

By contracting $\nu_1$ and $\lambda_3$ in \eqref{Qo3}, and antisymmetrizing $\nu_2,\dots,\nu_p$, and $\lambda_1$,
we obtain
\bea
0 & = & (D-p+4)(S_{\mu_1\dots\mu_p}{}^{\nu_1\dots\nu_p\lambda} + S_{\mu_1\dots\mu_p}{}^{\lambda[\nu_2\dots\nu_p\nu_1]})
\nn & & 
-pS^\lambda{}_{[\mu_2\dots\mu_p}{}^{\nu_1\dots\nu_p}{}_{\mu_1]}
-pS^{[\nu_1}{}_{[\mu_2\dots\mu_p}{}^{|\lambda|\nu_2\dots\nu_p]}{}_{\mu_1]}
\nn & &
-pS^{[\nu_1}{}_{[\mu_2\dots\mu_p\mu_1]}{}^{\nu_2\dots\nu_p]\lambda}
-pS^\lambda{}_{[\mu_2\dots\mu_p\mu_1]}{}^{[\nu_2\dots\nu_p\nu_1]}
\nn & &
+\text{(terms proportional to $S_{\alpha_1\dots\alpha_p}{}^{\beta_1\dots\beta_p\gamma}$
 with some pairs of indices contracted)}.
\label{Qo3-3}
\eea
By symmetrizing $\mu_1$, $\nu_1$, and $\lambda$ in \eqref{Qo3-3},
\bea
0 & = & \Big(D-p+3-\frac{2}{p+1}\Big)S^{(\lambda_1}{}_{\mu_1\dots\mu_{p-1}}{}^{\lambda_2|\nu_1\dots\nu_{p-1}|\lambda_3)}
\nn & &
 -\frac{(p-1)^2}{p+1}S^{(\lambda_1|\nu_1|}{}_{[\mu_2\dots\mu_{p-1}}{}^{\lambda_2}{}_{\mu_1]}{}^{|\nu_2\dots\nu_{p-1}|\lambda_3)}
\nn & &
+\text{(terms proportional to $S_{\alpha_1\dots\alpha_p}{}^{\beta_1\dots\beta_p\gamma}$
 with some pairs of indices contracted)}.
\label{Qo3-3-1}
\eea
In this equation it is understood that indices $\nu_1,\dots,\nu_p$ are antisymmetrized.
We can show that the following holds for $1\le n\le p-1$:
\bea
\lefteqn{S^{\lambda_1[\nu_1\dots\nu_n}{}_{[\mu_{n+1}\dots\mu_{p-1}}{}^{|\lambda_2|}{}_{\mu_1\dots\mu_n]}
{}^{\nu_{n+1}\dots\nu_{p-1}]\lambda_3}
} \nn & = & 
a_nS^{\lambda_1}{}_{\mu_1\dots\mu_{p-1}}{}^{\lambda_2\nu_1\dots\nu_{p-1}\lambda_3}
\nn & &
+\text{(terms proportional to $S_{\alpha_1\dots\alpha_p}{}^{\beta_1\dots\beta_p\gamma}$
 with some pairs of indices contracted)},
\label{Qo3-3-1-1}
\eea
where $a_n$ are constants, and in this equation it is understood that indices $\lambda_1,\lambda_2$, and $\lambda_3$
are symmetrized. Indeed, \eqref{Qo3-3-1-1} for $n=1$ is \eqref{Qo3-3-1} itself with
\beq
a_1=\frac{p+1}{(p-1)^2}\Big(D-p+3-\frac{2}{p+1}\Big).
\label{a1forpn1}
\eeq
By renaming $\mu_1$ and $\nu_1$ to $\nu_1$ and $\mu_1$ respectively, and antisymmetrizing $\mu_1,\dots,\mu_{p-1}$
and $\nu_1,\dots,\nu_{p-1}$ in \eqref{Qo3-3-1-1}, we obtain
\bea
\lefteqn{
a_nS^{\lambda_1[\nu_1}{}_{[\mu_2\dots\mu_{p-1}}{}^{|\lambda_2|}{}_{\mu_1]}{}^{\nu_2\dots\nu_{p-1}]\lambda_3}
} \nn & = & 
 \frac{n^2}{(p-1)^2}
 S^{\lambda_1[\nu_1\dots\nu_{n-1}}{}_{[\mu_n\dots\mu_{p-1}}{}^{|\lambda_2|}{}_{\mu_1\dots\mu_{n-1}]}
{}^{\nu_n\dots\nu_{p-1}]\lambda_3}
\nn & &
 -\frac{2n(p-n-1)}{(p-1)^2}
 S^{\lambda_1[\nu_1\dots\nu_n}{}_{[\mu_{n+1}\dots\mu_{p-1}}{}^{|\lambda_2|}{}_{\mu_1\dots\mu_n]}
{}^{\nu_{n+1}\dots\nu_{p-1}]\lambda_3}
\nn & &
 + \frac{(p-n-1)^2}{(p-1)^2}
 S^{\lambda_1[\nu_1\dots\nu_{n+1}}{}_{[\mu_{n+2}\dots\mu_{p-1}}{}^{|\lambda_2|}{}_{\mu_1\dots\mu_{n+1}]}
{}^{\nu_{n+2}\dots\nu_{p-1}]\lambda_3}
\nn & &
+\text{(terms proportional to $S_{\alpha_1\dots\alpha_p}{}^{\beta_1\dots\beta_p\gamma}$
 with some pairs of indices contracted)},
\eea
where it is understood that indices $\lambda_1,\lambda_2$, and $\lambda_3$ are symmetrized.
Therefore we obtain \eqref{Qo3-3-1-1} with $n$ replaced by $n+1$, with
\beq
\frac{n^2}{(p-1)^2}a_{n-1} - \frac{2n(p-n-1)}{(p-1)^2}a_n + \frac{(p-n-1)^2}{(p-1)^2}a_{n+1}=a_1a_n.
\label{Qo3-an}
\eeq
This equation, $a_1$ and $a_0=1$ uniquely determine $a_n$.
Then for $n=p-1$,
\bea
\lefteqn{
S^{(\lambda_1|\nu_1\dots\nu_{p-1}|\lambda_2}{}_{\mu_1\dots\mu_{p-1}}{}^{\lambda_3)}
} \nn & = & 
a_{p-1}S^{(\lambda_1}{}_{\mu_1\dots\mu_{p-1}}{}^{\lambda_2|\nu_1\dots\nu_{p-1}|\lambda_3)}
\nn & & 
+\text{(terms proportional to $S_{\alpha_1\dots\alpha_p}{}^{\beta_1\dots\beta_p\gamma}$
 with some pairs of indices contracted)},
\eea
and by interchanging $\mu_1,\dots,\mu_{p-1}$ and $\nu_1,\dots,\nu_{p-1}$,
\bea
\lefteqn{
S^{(\lambda_1}{}_{\mu_1\dots\mu_{p-1}}{}^{\lambda_2|\nu_1\dots\nu_{p-1}|\lambda_3)}
} \nn & = & 
a_{p-1}S^{(\lambda_1|\nu_1\dots\nu_{p-1}|\lambda_2}{}_{\mu_1\dots\mu_{p-1}}{}^{\lambda_3)}
\nn & & 
+\text{(terms proportional to $S_{\alpha_1\dots\alpha_p}{}^{\beta_1\dots\beta_p\gamma}$
 with some pairs of indices contracted)}.
\eea
If $a_{p-1}\neq\pm 1$, we can solve the above two equations and obtain
\bea
\lefteqn{
S^{(\lambda_1}{}_{\mu_1\dots\mu_{p-1}}{}^{\lambda_2|\nu_1\dots\nu_{p-1}|\lambda_3)}
} \nn & = & 
\text{(terms proportional to $S_{\alpha_1\dots\alpha_p}{}^{\beta_1\dots\beta_p\gamma}$
 with some pairs of indices contracted)}.
\label{Qo3-3-4}
\eea
$a_{p-1}\neq\pm 1$ can be explicitly confirmed for lower $p$:
\bea
\bullet~p=2 & & a_1 = 3D+1>1 \nn
\bullet~p=3 & & a_2 = 4\Big(D-\frac{1}{4}\Big)^2-\frac{5}{4}>1 \nn
\bullet~p=4 & & a_3 = 1+\frac{5}{36}\Big(D-\frac{7}{5}\Big)\Big[25\Big(D-\frac{3}{5}\Big)^2-7\Big]>1
\nonumber
\eea
For $p=1$, \eqref{a1forpn1} does not make sense, but in this case \eqref{Qo3-3-1} directly
gives \eqref{Qo3-3-4}.

By symmetrizing $\nu_1$ and $\lambda$ in \eqref{Qo3-3},
\bea
0 & = &
 \frac{p+1}{p}(D+4)S_{\mu_1\dots\mu_p}{}^{(\lambda_1|\nu_1\dots\nu_{p-1}|\lambda_2)}
\nn & &
 +(p-1)(S^{(\lambda_1}{}_{[\mu_2\dots\mu_p\mu_1]}{}^{|\nu_1\dots\nu_{p-1}|\lambda_2)}
 +S^{\nu_1}{}_{[\mu_2\dots\mu_p\mu_1]}{}^{(\lambda_1|\nu_2\dots\nu_{p-1}|\lambda_2)}
 +S^{(\lambda_1}{}_{[\mu_2\dots\mu_p\mu_1]}{}^{\lambda_2)|\nu_2\dots\nu_{p-1}\nu_1}
)
\nn & & 
 -(p+1)(
 S^{(\lambda_1}{}_{[\mu_2\dots\mu_p}{}^{\lambda_2)\nu_1\dots\nu_{p-1}}{}_{\mu_1]}
 +S_{\mu_1\dots\mu_p}{}^{(\lambda_1|\nu_1\dots\nu_{p-1}|\lambda_2)}
 +S^{(\lambda_1}{}_{[\mu_2\dots\mu_p\mu_1]}{}^{|\nu_1\dots\nu_{p-1}|\lambda_2)}
)
\nn & & 
+\text{(terms proportional to $S_{\alpha_1\dots\alpha_p}{}^{\beta_1\dots\beta_p\gamma}$
 with some pairs of indices contracted)}.
\eea
In this equation it is understood that indices $\nu_1,\dots,\nu_{p-1}$ are antisymmetrized.
The expressions in the second and third lines in this equation are symmetric under the interchange of
the first, $(p+1)$-th, and $(2p+1)$-th indices,
and therefore by \eqref{Qo3-3-4},
\bea
\lefteqn{S_{\mu_1\dots\mu_p}{}^{(\lambda_1|\nu_1\dots\nu_{p-1}|\lambda_2)}
} \nn & = & 
\text{(terms proportional to $S_{\alpha_1\dots\alpha_p}{}^{\beta_1\dots\beta_p\gamma}$
 with some pairs of indices contracted)}.
\label{Qo3-3-4-1}
\eea

By contracting $\mu_1$ and $\lambda_3$ in \eqref{Qo3},
\bea
0 & = & \Big[ D-p+3-\frac{2}{D+2} -\frac{p}{D-p+1}\Big(3-\frac{2p}{D+2}\Big) \Big]
 S^{(\lambda_1}{}_{\mu_1\dots\mu_{p-1}}{}^{|\nu_1\dots\nu_p|\lambda_2)}
\nn & &
 +\frac{p(p-1)}{D-p+1}\frac{D+4}{D+2}
  S^{(\lambda_1}{}_{[\mu_1\dots\mu_{p-2}}{}^{|\nu_1}{}_{\mu_{p-1}]}{}^{\nu_2\dots\nu_p|\lambda_2)}
\nn & &
 -\frac{p}{D-p+1}(D+4)S^{\nu_1}{}_{\mu_1\dots\mu_{p-1}}{}^{(\lambda_1|\nu_2\dots\nu_p|\lambda_2)}
\nn & &
 +\frac{p}{D-p+1}\Big( 3-\frac{2p}{D+2} \Big)\Big[
 S^{(\lambda_1}{}_{\mu_1\dots\mu_{p-1}}{}^{|\nu_1\dots\nu_p|\lambda_2)}
 +S^{(\lambda_1}{}_{\mu_1\dots\mu_{p-1}}{}^{\lambda_2)\nu_2\dots\nu_p\nu_1}
\nn & &
 +S^{\nu_1}{}_{\mu_1\dots\mu_{p-1}}{}^{(\lambda_1|\nu_2\dots\nu_p|\lambda_2)}
\Big]
\nn & &
 +\frac{p(p-1)}{D-p+1}\frac{D+4}{D+2}\Big[
 S^{\nu_1}{}_{[\mu_1\dots\mu_{p-2}}{}^{(\lambda_1\lambda_2)}{}^{\nu_2\dots\nu_p}{}_{\mu_{p-1}]}
\nn & &
 +S^{\nu_1}{}_{\mu_1\dots\mu_{p-1}}{}^{(\lambda_1|\nu_2\dots\nu_p|\lambda_2)}
 +S^{\nu_1}{}_{[\mu_1\dots\mu_{p-2}}{}^{(\lambda_1|}{}_{\mu_{p-1}]}{}^{\nu_2\dots\nu_p|\lambda_2)}
\Big]
\nn & & 
+\text{(terms proportional to $S_{\alpha_1\dots\alpha_p}{}^{\beta_1\dots\beta_p\gamma}$
 with some pairs of indices contracted)},
\label{Qo3-aux1}
\eea
where it is understood that indices $\nu_1,\dots,\nu_p$ are antisymmetrized.
The expression in the fourth and fifth lines of the above is symmetric under the interchange of 
the first, $(p+1)$-th and $(2p+1)$-th indices, and the expression in the sixth and seventh lines
is symmetric under the interchange of the $p$-th, $(p+1)$-th and $(2p+1)$-th indices.
Therefore \eqref{Qo3-3-4} can be applied to them. 
\eqref{Qo3-3-4-1} can also be applied to the third term. Then we obtain
\bea
\lefteqn{
S^{(\lambda_1}{}_{[\mu_1\dots\mu_{p-2}}{}^{|[\nu_1}{}_{\mu_{p-1}]}{}^{\nu_2\dots\nu_p]|\lambda_2)}
} \nn & = & b_1S^{(\lambda_1}{}_{\mu_1\dots\mu_{p-1}}{}^{|\nu_1\dots\nu_p|\lambda_2)}
\nn & & 
+\text{(terms proportional to $S_{\alpha_1\dots\alpha_p}{}^{\beta_1\dots\beta_p\gamma}$
 with some pairs of indices contracted)},
\eea
where
\beq
b_1=-\frac{D-p+1}{p(p-1)}\frac{D+2}{D+4}\Big[ D-p+3-\frac{2}{D+2} -\frac{p}{D-p+1}\Big(3-\frac{2p}{D+2}\Big) \Big].
\eeq
Similarly to \eqref{Qo3-3-1}-\eqref{Qo3-an},
we can show that the following holds for $1\le n\le p-1$:
\bea
\lefteqn{
S^{(\lambda_1}{}_{[\mu_1\dots\mu_{p-n-1}}{}^{|[\nu_1\dots\nu_n}{}_{\mu_{p-n}\dots\mu_{p-1}]}{}^{\nu_{n+1}\dots\nu_p]|\lambda_2)}
} \nn & = & b_nS^{(\lambda_1}{}_{\mu_1\dots\mu_{p-1}}{}^{|\nu_1\dots\nu_p|\lambda_2)}
\nn & & 
+\text{(terms proportional to $S_{\alpha_1\dots\alpha_p}{}^{\beta_1\dots\beta_p\gamma}$
 with some pairs of indices contracted)},
\eea
where $b_n$ satisfy
\beq
\frac{n^2}{p(p-1)}b_{n-1} - \frac{n(2p-2n-1)}{p(p-1)}b_n + \frac{(p-n-1)(p-n)}{p(p-1)}b_{n+1}=b_1b_n.
\eeq
This equation, $b_0=1$ and $b_1$ determine $b_n$ uniquely.
Then for $n=p-1$,
\bea
\lefteqn{
S^{(\lambda_1|[\nu_1\dots\nu_{p-1}}{}_{\mu_1\dots\mu_{p-1}}{}^{\nu_p]|\lambda_2)}
} \nn & = & b_{p-1}S^{(\lambda_1}{}_{\mu_1\dots\mu_{p-1}}{}^{|\nu_1\dots\nu_p|\lambda_2)}
\nn & & 
+\text{(terms proportional to $S_{\alpha_1\dots\alpha_p}{}^{\beta_1\dots\beta_p\gamma}$
 with some pairs of indices contracted)},
\eea
and by interchanging $\mu_1,\dots,\mu_{p-1}$ and $\nu_1,\dots,\nu_{p-1}$, and
antisymmetrizing $\nu_1,\dots,\nu_p$ in this equation,
\bea
\lefteqn{
S^{(\lambda_1}{}_{\mu_1\dots\mu_{p-1}}{}^{|\nu_1\dots\nu_p|\lambda_2)}
 - (p-1)S^{(\lambda_1}{}_{[\mu_1\dots\mu_{p-2}}{}^{|[\nu_p\nu_1\dots\nu_{p-1}]|}{}_{\mu_{p-1}]}{}^{\lambda_2)}
} \nn & = & pb_{p-1}S^{(\lambda_1|[\nu_1\dots\nu_{p-1}}{}_{\mu_1\dots\mu_{p-1}}{}^{\nu_p]|\lambda_2)}
\nn & & 
+\text{(terms proportional to $S_{\alpha_1\dots\alpha_p}{}^{\beta_1\dots\beta_p\gamma}$
 with some pairs of indices contracted)}.
\eea
Therefore,
\bea
\lefteqn{
[pb_{p-1}^2+(p-1)b_1-1]S^{(\lambda_1}{}_{\mu_1\dots\mu_{p-1}}{}^{|\nu_1\dots\nu_p|\lambda_2)}
} \nn & = & 
\text{(terms proportional to $S_{\alpha_1\dots\alpha_p}{}^{\beta_1\dots\beta_p\gamma}$
 with some pairs of indices contracted)}.
\eea
If $pb_{p-1}^2+(p-1)b_1-1\neq 0$, we obtain
\bea
\lefteqn{
S^{(\lambda_1}{}_{\mu_1\dots\mu_{p-1}}{}^{|\nu_1\dots\nu_p|\lambda_2)}
} \nn & = & 
\text{(terms proportional to $S_{\alpha_1\dots\alpha_p}{}^{\beta_1\dots\beta_p\gamma}$
 with some pairs of indices contracted)}.
\label{Qo3-3-5}
\eea
For lower $p$, $pb_{p-1}^2+(p-1)b_1-1$ is equal to the followings:
\bea
\bullet~ p=2 &  & \frac{1}{2}(D-3)(D-2)D(D+1) \nn
\bullet~ p=3 &  & \frac{1}{48}(D-5)(D-4)^2(D-3)(D-1)D^2(D+1) \nn
\bullet~ p=4 &  & \frac{1}{5184}(D-7)(D-6)(D-5)(D-4)(D-2)(D-1)D(D+1) \nn
& & \times(D^2-8D+9) (D^2-4D-3)
\nonumber
\eea
Therefore, for example $(p,D)=(2,3), (3,4)$ and $(3,5)$ we need special care, but here we 
assume $pb_{p-1}^2+(p-1)b_1-1\neq 0$ and proceed with the analysis.
For $p=1$, \eqref{Qo3-aux1} directly gives \eqref{Qo3-3-5}.

Then from \eqref{Qo3-3-4}, \eqref{Qo3-3-4-1}, and \eqref{Qo3-3-5}, 
\bea
\lefteqn{
S^{(\lambda_1}{}_{\mu_1\dots\mu_{p-1}}{}^{\lambda_2)\nu_1\dots\nu_{p-1}\lambda_3}
} \nn & = & 
 3S^{(\lambda_1}{}_{\mu_1\dots\mu_{p-1}}{}^{\lambda_2|\nu_1\dots\nu_{p-1}|\lambda_3)}
 - S^{\lambda_3}{}_{\mu_1\dots\mu_{p-1}}{}^{(\lambda_1|\nu_1\dots\nu_{p-1}|\lambda_2)}
 - S^{(\lambda_1}{}_{\mu_1\dots\mu_{p-1}}{}^{|\lambda_3\nu_1\dots\nu_{p-1}|\lambda_2)}
\nn & = & 
\text{(terms proportional to $S_{\alpha_1\dots\alpha_p}{}^{\beta_1\dots\beta_p\gamma}$
 with some pairs of indices contracted)}.
\label{Qo3-3-6}
\eea
\eqref{Qo3-3-4-1}, and \eqref{Qo3-3-5}, and \eqref{Qo3-3-6} mean that
$S_{\mu_1\dots\mu_p}{}^{\nu_1\dots\nu_p\lambda}$ with two indices exchanged is equal to
$-S_{\mu_1\dots\mu_p}{}^{\nu_1\dots\nu_p\lambda}$ plus extra terms with some pairs of indices contracted.
Then if we define a $(2p+1)$-form $Y_{(2p+1)\mu_1\dots\mu_p\nu_1\dots\nu_p\lambda}$ as 
$Y_{(2p+1)\mu_1\dots\mu_p\nu_1\dots\nu_p\lambda}=S_{[\mu_1\dots\mu_p\nu_1\dots\nu_p\lambda]}$,
the difference between $S_{\mu_1\dots\mu_p}{}^{\nu_1\dots\nu_p\lambda}$ and
 $Y_{(2p+1)\mu_1\dots\mu_p}{}^{\nu_1\dots\nu_p\lambda}$ consists of terms with some pairs of indices contracted:
\bea
S_{\mu_1\dots\mu_p}{}^{\nu_1\dots\nu_p\lambda}
 & = & Y_{(2p+1)\mu_1\dots\mu_p}{}^{\nu_1\dots\nu_p\lambda}
\nn & & 
+ \text{(terms proportional to $S_{\alpha_1\dots\alpha_p}{}^{\beta_1\dots\beta_p\gamma}$}
\nn & & 
 \text{with some pairs of indices contracted)}.
\eea
For example, for $p=1$ this can be explicitly seen by the following: 
\bea
Y_{(3)\mu\nu\lambda} & = & S_{[\mu\nu\lambda]}
\nn & = & 
 S_{\mu\nu\lambda} + \frac{1}{3}( S_{(\mu|\lambda|\nu)} - S_{(\mu\nu)\lambda} + S_{\lambda(\mu\nu)} 
- 2S_{(\mu|\nu|\lambda)} - 2S_{\mu(\nu\lambda)} ).
\eea

Thus $S_{\mu_1\dots\mu_p}{}^{\nu_1\dots\nu_p\lambda}$ is expressed by 
$Y_{(2p+1)\mu_1\dots\mu_p}{}^{\nu_1\dots\nu_p\lambda}$ and tensors with less free indices.
Further analysis of \eqref{Qo3} reveals the structure of those tensors. However it is difficult to 
continue such analysis for arbitrary $p$. So we will only give results of the analysis for lower $p$ in the next section,
and proceed to the next step here.

\paragraph{Analysis of \eqref{qo2}}

By contracting $\lambda_1$ and $\lambda_2$ in \eqref{qo2},
\bea
q_{\mu_1\dots\mu_p}{}^{\nu_1\dots\nu_p} & = & s_{\mu_1\dots\mu_p}{}^{\nu_1\dots\nu_p}
\nn & & +\frac{1}{D-p}(
 2\nabla_\rho S_{\mu_1\dots\mu_p}{}^{\nu_1\dots\nu_p\rho}
 -p\nabla_\rho S^\rho{}_{[\mu_2\dots\mu_p}{}^{\nu_1\dots\nu_p}{}_{\mu_1]}
\nn & & 
 -p\nabla_{[\mu_1}S^\rho{}_{\mu_2\dots\mu_p]}{}^{\nu_1\dots\nu_p}{}_\rho
).
\label{qo2-q0}
\eea
Then we can eliminate $q_{\mu_1\dots\mu_p}{}^{\nu_1\dots\nu_p}$ from \eqref{qo2}.
By contracting $\mu_1$ and $\lambda_2$ in \eqref{qo2}, 
\bea
(D-p+1)s^\lambda{}_{\mu_1\dots\mu_{p-1}}{}^{\nu_1\dots\nu_p}
+(p+1)s^{[\lambda}{}_{\mu_1\dots\mu_{p-1}}{}^{\nu_1\dots\nu_p]}
-ps_{\rho\mu_1\dots\mu_{p-1}}{}^{\rho[\nu_2\dots\nu_p}g^{\nu_1]\lambda}
 \nn  =  
\text{(terms proportional to $\nabla_\delta S_{\alpha_1\dots\alpha_p}{}^{\beta_1\dots\beta_p\gamma}$)}.
\label{qo2-aux1}
\eea
By antisymmetrizing $\lambda, \nu_1,\dots,\nu_p$ in the above,
\beq
s^{[\lambda}{}_{\mu_1\dots\mu_{p-1}}{}^{\nu_1\dots\nu_p]} = 
\text{(terms proportional to $\nabla_\delta S_{\alpha_1\dots\alpha_p}{}^{\beta_1\dots\beta_p\gamma}$)}.
\eeq
Then from the above two,
\bea
s^\lambda{}_{\mu_1\dots\mu_{p-1}}{}^{\nu_1\dots\nu_p}
& = &
\frac{p}{D-p+1}s_{\rho\mu_1\dots\mu_{p-1}}{}^{\rho[\nu_2\dots\nu_p}g^{\nu_1]\lambda}
\nn & & 
+\text{(terms proportional to $\nabla_\delta S_{\alpha_1\dots\alpha_p}{}^{\beta_1\dots\beta_p\gamma}$)}.
\eea
By antisymmetrizing $\lambda, \mu_1,\dots,\mu_{p-1}$ in the above,
\bea
s_{\mu_1\dots\mu_p}{}^{\nu_1\dots\nu_p}
& = &
\frac{p}{D-p+1}s_{\rho[\mu_2\dots\mu_p}{}^{\rho[\nu_2\dots\nu_p}\delta_{\mu_1]}{}^{\nu_1]}
\nn & & 
+\text{(terms proportional to $\nabla_\delta S_{\alpha_1\dots\alpha_p}{}^{\beta_1\dots\beta_p\gamma}$)}.
\label{qo2-aux2}
\eea
From this we can show the following for $1\le n\le p$:
\bea
s_{\mu_1\dots\mu_p}{}^{\nu_1\dots\nu_p}
& = &
d_ns_{\rho_1\dots\rho_n[\mu_1\dots\mu_{p-n}}{}^{\rho_1\dots\rho_n[\nu_1\dots\nu_{p-n}}
 \delta_{\mu_{p-n+1}}{}^{\nu_{p-n+1}}\dots\delta_{\mu_p]}{}^{\nu_p]}
\nn & & 
+\text{(terms proportional to $\nabla_\delta S_{\alpha_1\dots\alpha_p}{}^{\beta_1\dots\beta_p\gamma}$)},
\label{qo2-aux3}
\eea
where $d_n$ are constants. Indeed \eqref{qo2-aux3} for $n=1$ is \eqref{qo2-aux2} itself with 
\beq
d_1=\frac{p}{D-p+1},
\eeq
and by contracting $\mu_p$ and $\nu_p(=\rho_{n+1})$, putting $\delta_{\mu_p}{}^{\nu_p}$ on both sides, and
antisymmetrizing $\mu_1,\dots,\mu_p$ and $\nu_1,\dots,\nu_p$ in \eqref{qo2-aux3}, we find that
\beq
d_{n+1}=\frac{p-n}{D-p+n+1}d_n.
\eeq
Then $d_p=\frac{p!(D-p)!}{D!}$ and 
\bea
s_{\mu_1\dots\mu_p}{}^{\nu_1\dots\nu_p}
& = &
\frac{p!(D-p)!}{D!}s_{\rho_1\dots\rho_p}{}^{\rho_1\dots\rho_p}
 \delta_{[\mu_1}{}^{\nu_1}\dots\delta_{\mu_p]}{}^{\nu_p}
\nn & & 
+\text{(terms proportional to $\nabla_\delta S_{\alpha_1\dots\alpha_p}{}^{\beta_1\dots\beta_p\gamma}$)}.
\label{qo2-s}
\eea
From this and \eqref{qo2-q0},
\bea
q_{\mu_1\dots\mu_p}{}^{\nu_1\dots\nu_p}
& = &
\frac{p!(D-p)!}{D!}s_{\rho_1\dots\rho_p}{}^{\rho_1\dots\rho_p}
 \delta_{[\mu_1}{}^{\nu_1}\dots\delta_{\mu_p]}{}^{\nu_p}
\nn & & 
+\text{(terms proportional to $\nabla_\delta S_{\alpha_1\dots\alpha_p}{}^{\beta_1\dots\beta_p\gamma}$)}.
\label{qo2-q}
\eea
By \eqref{qo2-s} and \eqref{qo2-q}, we obtain an equation consisting of only 
$\nabla_\delta S_{\alpha_1\dots\alpha_p}{}^{\beta_1\dots\beta_p\gamma}$ from \eqref{qo2}.
Therefore \eqref{qo2} contains no more information on $s_{\mu_1\dots\mu_p}{}^{\nu_1\dots\nu_p}$ and
$q_{\mu_1\dots\mu_p}{}^{\nu_1\dots\nu_p}$.

The solution to \eqref{Qo3} gives the details of $s_{\mu_1\dots\mu_p}{}^{\nu_1\dots\nu_p}$ and
$q_{\mu_1\dots\mu_p}{}^{\nu_1\dots\nu_p}$, and we can obtain further information on the tensors in the solution 
from \eqref{qo2}. However such analysis for arbitrary $p$ is complicated, and therefore we give results only for lower $p$
in the next section.

\section{Results for Lower $p$}

In this section we show the results of the analyses outlined in the previous section for lower $p$,
in order to obtain some insight into the structure of the solutions to the conditions for arbitrary $p$.
For $p=1$ and $p=2$ we give the general solutions to the conditions \eqref{Qo3}-\eqref{O0}.
For $p=3$ we show the general solution to only \eqref{Qo3} and \eqref{qo2}.

Although we do not explain full details of the derivation process of the results because they are very 
lengthy and essentially the repetition of analyses similar to the one in the previous section, 
we can give comments on the features common to these lower $p$ cases:
As has been explained in the previous section,
the structures of  $S_{\mu_1\dots\mu_p}{}^{\nu_1\dots\nu_p\lambda}$
and $Q_{\mu_1\dots\mu_p}{}^{\nu_1\dots\nu_p\lambda}$ are determined by solving \eqref{Qo3}.
Then by solving \eqref{qo2}, the structures of $s_{\mu_1\dots\mu_p}{}^{\nu_1\dots\nu_p}$ and 
$q_{\mu_1\dots\mu_p}{}^{\nu_1\dots\nu_p}$ are determined, and it turns out that
they consist of Killing-Yano forms, or conformal Killing Yano forms, etc.
Then \eqref{O1} and \eqref{O0} determine more on especially $c$ and $K^\mu$ (see below) with no need of 
the background equation of motion.

\paragraph{Results for $p=1$}
\
In this case we consider $D\ge 2$, and 
first we give the general solution to \eqref{Qo3}-\eqref{O0} for the massless case $m=0$.
It consists of  $c$, $U^\nu$, $V_\mu$, $K^\mu$, and $Y_{\mu_1\mu_2\mu_3}$, where
$c$ is a constant, and $U^\nu$ and $V_\mu$ are arbitrary vectors.
The rest depend on $D$:
\begin{alignat}{2}
 \bullet~D &=2 & \quad & \text{$K^\mu$ is a vector satisfying $\nabla_\rho K^\rho=$ const. and $Y_{\mu_1\mu_2\mu_3}=0$.}
\notag\\
 \bullet~D &=4 & & \text{$K^\mu$ is a conformal Killing vector and $Y_{\mu_1\mu_2\mu_3}$ is a conformal Killing-Yano form.}
\notag\\
 \bullet~D &\neq 2,4 & & \text{$K^\mu$ is a conformal Killing vector with $\nabla_\rho K^\rho=$ const.}
\notag\\
 & & & \text{and $Y_{\mu_1\mu_2\mu_3}$ is a Killing-Yano form.}
\notag
\end{alignat}
In either case, $S_\mu{}^{\nu\lambda}$, $Q_\mu{}^{\nu\lambda}$ and $s_\mu{}^\nu$ are given as follows:
\bea
S_\mu{}^{\nu\lambda} & = &
 \delta_\mu{}^\lambda U^\nu + \delta_\mu{}^\nu K^\lambda + Y_\mu{}^{\nu\lambda},
\\
Q_\mu{}^{\nu\lambda} & = & 
 g^{\lambda\nu}V_\mu + \delta_\mu{}^\nu K^\lambda + Y_\mu{}^{\nu\lambda},
\\
s_\mu{}^\nu & = & c\delta_\mu{}^\nu + \nabla_\mu U^\nu + \nabla_\mu K^\nu,
\eea
and $q_\mu{}^\nu$ depends on $D$:
\begin{alignat}{2}
\bullet~D &\neq 2,4 & \quad
 q_\mu{}^\nu & = \Big(c+\frac{2}{D}\nabla_\rho K^\rho\Big)\delta_\mu{}^\nu + \nabla_\mu K^\nu.
\\
\bullet~D &=2 & 
 q_\mu{}^\nu &= \Big(c+2\nabla_\rho K^\rho\Big)\delta_\mu{}^\nu - \nabla^\nu K_\mu.
\\
\bullet~D &=4 & 
 q_\mu{}^\nu &= \Big(c+\frac{1}{2}\nabla_\rho K^\rho\Big)\delta_\mu{}^\nu + \nabla_\mu K^\nu
 + \nabla_\rho Y^\rho{}_\mu{}^\nu.
\end{alignat}

Next we give the general solution to \eqref{Qo3}-\eqref{O0} for the massive case $m\neq 0$.
It consists of $c$, $K^\mu$, and $Y_{\mu_1\mu_2\mu_3}$, where
$Y_{\mu_1\mu_2\mu_3}$ is a Killing-Yano form, $c$ and $K^\mu$ satisfy
\bea
\nabla_\mu\nabla_\nu c & = & \frac{1}{D}g_{\mu\nu}\nabla^2 c, 
\label{p2mn0-1} \\
m^2\nabla_\rho K^\rho & = & -\frac{1}{2}(D-2)\nabla^2c, 
\label{p2mn0-2} \\
0 & = & \nabla_\mu\Big(c-\frac{D-4}{2D}\nabla_\rho K^\rho\Big),
\label{p2mn0-3}
\eea
and then
\begin{alignat}{2}
\bullet~D &=2 & \quad & \text{$c$ is a constant, $K^\mu$ is a Killing vector, and $Y_{\mu_1\mu_2\mu_3}=0$.}
\notag\\
\bullet~D &=4 & \quad & \text{$c$ is a constant, and $K^\mu$ is a Killing vector.}
\notag\\
\bullet~D &\neq 2,4 & \quad & \text{$c$ is a function and $K^\mu$ is a conformal Killing vector satisfying
 \eqref{p2mn0-1}-\eqref{p2mn0-3}.}
\notag
\end{alignat}
In either case,
\bea
S_\mu{}^{\nu\lambda} & = &
 \frac{1}{m^2}\delta_\mu{}^\lambda \nabla^\nu c + \delta_\mu{}^\nu K^\lambda + Y_\mu{}^{\nu\lambda},
\\
Q_\mu{}^{\nu\lambda} & = & 
 \frac{1}{m^2}g^{\lambda\nu}\nabla_\mu c + \delta_\mu{}^\nu K^\lambda + Y_\mu{}^{\nu\lambda},
\\
s_\mu{}^\nu & = & \Big(c+\frac{1}{Dm^2}\nabla^2c\Big)\delta_\mu{}^\nu + \nabla_\mu K^\nu,
\\
q_\mu{}^\nu & = & \Big(c-\frac{D-2}{D}\frac{1}{m^2}\nabla^2c
\Big)\delta_\mu{}^\nu + \nabla_\mu K^\nu.
\eea

\paragraph{Results for $p=2$}
\
In this case we consider $D\ge 3$, and 
first we give the general solution to \eqref{Qo3}-\eqref{O0} for the massless case $m=0$.
It consists of  
$c$, $U_{\mu}{}^{\nu_1\nu_2}$, $V_{\mu_1\mu_2}{}^\nu$, $K^\mu$, $Y_{(3)\mu_1\mu_2\mu_3}$ and $Y_{(5)\mu_1\dots\mu_5}$,
where
$c$ is a constant, and $U_\mu{}^{\nu_1\nu_2}$ and $V_{\mu_1\mu_2}{}^\nu$ are arbitrary tensors.
The rest depends on $D$:
\begin{alignat}{2}
\bullet~D & =3 & \quad & \text{$K^\mu$ is a vector satisfying $\nabla_\rho K^\rho=$ const.,}
\notag\\ & & &
  \text{$Y_{(3)\mu_1\mu_2\mu_3}=0$, and $Y_{(5)\mu_1\dots\mu_5}=0$.}
\notag\\
\bullet~D & =4 & \quad & \text{$K^\mu$ is a conformal Killing vector with $\nabla_\rho K^\rho=$ const.,}
\notag\\ & & &
  \text{$Y_{(3)\mu_1\mu_2\mu_3}$ is a coclosed 3-form, and $Y_{(5)\mu_1\dots\mu_5}=0$.}
\notag\\
\bullet~D & =6 & \quad & \text{$K^\mu$ is a conformal Killing vector, and}
\notag\\ & & &
  \text{$Y_{(3)\mu_1\mu_2\mu_3}$ and $Y_{(5)\mu_1\dots\mu_5}$ are conformal Killing-Yano forms.}
\notag\\
\bullet~D &\neq 3,4,6 & \quad & \text{$K^\mu$ is a conformal Killing vector with $\nabla_\rho K^\rho=$ const., and }
\notag\\ & & &
  \text{$Y_{(3)\mu_1\mu_2\mu_3}$ and $Y_{(5)\mu_1\dots\mu_5}$ are Killing-Yano forms.}
\notag
\end{alignat}
In either case,
\bea
S_{\mu_1\mu_2}{}^{\nu_1\nu_2\lambda} & = &
 \delta_{[\mu_1}{}^\lambda U_{\mu_2]}{}^{\nu_1\nu_2}
 + \delta_{[\mu_1}{}^{\nu_1}\delta_{\mu_2]}{}^{\nu_2}K^\lambda
 + \delta_{[\mu_1}{}^{[\nu_1}Y_{(3)\mu_2]}{}^{\nu_2]\lambda}
 + Y_{(5)\mu_1\mu_2}{}^{\nu_1\nu_2\lambda},
\\
Q_{\mu_1\mu_2}{}^{\nu_1\nu_2\lambda} & = & 
 g^{\lambda[\nu_1}V_{\mu_1\mu_2}{}^{\nu_2]}
 + \delta_{[\mu_1}{}^{\nu_1}\delta_{\mu_2]}{}^{\nu_2}K^\lambda
 + \delta_{[\mu_1}{}^{[\nu_1}Y_{(3)\mu_2]}{}^{\nu_2]\lambda}
 + Y_{(5)\mu_1\mu_2}{}^{\nu_1\nu_2\lambda},
\\
s_{\mu_1\mu_2}{}^{\nu_1\nu_2} & = &
 c\delta_{[\mu_1}{}^{\nu_1}\delta_{\mu_2]}{}^{\nu_2} + \nabla_{[\mu_1}U_{\mu_2]}{}^{\nu_1\nu_2}
\nn & & 
 + 2\delta_{[\mu_1}{}^{[\nu_1}\nabla_{\mu_2]}K^{\nu_2]}
 - \frac{1}{2}\nabla_{[\mu_1}Y_{(3)\mu_2]}{}^{\nu_1\nu_2}.
\eea
$q_{\mu_1\mu_2}{}^{\nu_1\nu_2}$ depends on $D$:
\begin{alignat}{2}
\bullet~D &\neq 3, 4, 6 & \quad
q_{\mu_1\mu_2}{}^{\nu_1\nu_2} & =
 \Big(c+\frac{2}{D}\nabla_\rho K^\rho\Big)\delta_{[\mu_1}{}^{\nu_1}\delta_{\mu_2]}{}^{\nu_2}
\notag\\ & & & \quad
 + 2\delta_{[\mu_1}{}^{[\nu_1}\nabla_{\mu_2]}K^{\nu_2]}
 - \frac{1}{2}\nabla_{[\mu_1}Y_{(3)\mu_2]}{}^{\nu_1\nu_2}.
\\
\bullet~D &=3 & \quad
q_{\mu_1\mu_2}{}^{\nu_1\nu_2} & =
 \Big(c+2\nabla_\rho K^\rho\Big)\delta_{[\mu_1}{}^{\nu_1}\delta_{\mu_2]}{}^{\nu_2} 
 - 2\delta_{[\mu_1}{}^{[\nu_1}\nabla^{\nu_2]}K_{\mu_2]}.
\\
\bullet~D &=4 & \quad
q_{\mu_1\mu_2}{}^{\nu_1\nu_2} & =
 \Big(c+\frac{1}{2}\nabla_\rho K^\rho\Big)\delta_{[\mu_1}{}^{\nu_1}\delta_{\mu_2]}{}^{\nu_2}
\notag\\ & & & \quad
 + 2\delta_{[\mu_1}{}^{[\nu_1}\nabla_{\mu_2]}K^{\nu_2]}
 - \frac{1}{2}\nabla^{[\nu_1}Y_{(3)}{}^{\nu_2]}{}_{\mu_1\mu_2}.
\\
\bullet~D &=6 & \quad
q_{\mu_1\mu_2}{}^{\nu_1\nu_2} & =
 \Big(c+\frac{1}{3}\nabla_\rho K^\rho\Big)\delta_{[\mu_1}{}^{\nu_1}\delta_{\mu_2]}{}^{\nu_2}
 + 2\delta_{[\mu_1}{}^{[\nu_1}\nabla_{\mu_2]}K^{\nu_2]}
\notag\\ & & & \quad
 - \frac{1}{2}\nabla_{[\mu_1}Y_{(3)\mu_2]}{}^{\nu_1\nu_2}
 + \frac{1}{2}\delta_{[\mu_1}{}^{[\nu_1}\nabla_{|\rho|}Y_{(3)\mu_2]}{}^{\nu_2]\rho}
 + \nabla_\rho Y_{(5)\mu_1\mu_2}{}^{\nu_1\nu_2\rho}.
\end{alignat}

Next we give the general solution to \eqref{Qo3}-\eqref{O0} for the massive case $m\neq 0$.
It consists of 
$c$, $u_{\mu_1\mu_2\mu_3}$, $K^\mu$, $Y_{(3)\mu_1\mu_2\mu_3}$ and $Y_{(5)\mu_1\dots\mu_5}$.
$c$ and $K^\mu$ satisfy
\bea
m^2\nabla_\rho K^\rho & = & -\frac{1}{2}(D-4)\nabla^2c, 
\label{p2cK1}\\
0 & = & \nabla_\mu\Big(c-\frac{D-6}{2D}\nabla_\rho K^\rho\Big),
\label{p2cK2}
\eea
and
\begin{alignat}{2}
\bullet~D &=3 & \quad & \text{$u_{\mu_1\mu_2\mu_3}$ is a coclosed 3-form,
 $K_\mu+\frac{1}{m^2}\nabla_\mu c$ is a conformal Killing vector,}
\notag\\ & & &
 \text{$Y_{(3)\mu_1\mu_2\mu_3}=0$, and $Y_{(5)\mu_1\dots\mu_5}=0$.}
\notag\\
\bullet~D &=4 & \quad & \text{$c$ is a constant, $u_{\mu_1\mu_2\mu_3}=0$, $K^\mu$ is a Killing vector,}
\notag\\ & & &
 \text{$Y_{(3)\mu_1\mu_2\mu_3}$ is a Killing-Yano form, and $Y_{(5)\mu_1\dots\mu_5}=0$.}
\notag\\
\bullet~D &=6 & \quad & \text{$c$ is a constant, $u_{\mu_1\mu_2\mu_3}=0$, $K^\mu$ is a Killing vector, and}
\notag\\ & & &
 \text{$Y_{(3)\mu_1\mu_2\mu_3}$ and $Y_{(5)\mu_1\dots\mu_5}$ are Killing-Yano forms.}
\notag\\
\bullet~D &\neq 3,4,6 & \quad & \text{$\nabla_\mu\nabla_\nu c = \frac{1}{D}g_{\mu\nu}\nabla^2 c$,
$u_{\mu_1\mu_2\mu_3}=0$,}
\notag\\ & & &
 \text{ $K^\mu$ is a conformal Killing vector satisfying \eqref{p2cK1} and \eqref{p2cK2}, and}
\notag\\ & & &
 \text{$Y_{(3)\mu_1\mu_2\mu_3}$ and $Y_{(5)\mu_1\dots\mu_5}$ are Killing-Yano forms.}
\notag
\end{alignat}
In either case,
\bea
S_{\mu_1\mu_2}{}^{\nu_1\nu_2\lambda} & = &
 -\frac{1}{m^2}\delta_{[\mu_1}{}^\lambda\delta_{\mu_2]}{}^{[\nu_1}\nabla^{\nu_2]} c
 + \delta_{[\mu_1}{}^\lambda u_{\mu_2]}{}^{\nu_1\nu_2}
\nn & &
 + \delta_{[\mu_1}{}^{\nu_1}\delta_{\mu_2]}{}^{\nu_2}K^\lambda
 + \delta_{[\mu_1}{}^{[\nu_1}Y_{(3)\mu_2]}{}^{\nu_2]\lambda}
 + Y_{(5)\mu_1\mu_2}{}^{\nu_1\nu_2\lambda},
\\
Q_{\mu_1\mu_2}{}^{\nu_1\nu_2\lambda} & = & 
 -\frac{1}{m^2}g^{\lambda[\nu_1}\delta_{[\mu_1}{}^{\nu_2]}\nabla_{\mu_2]} c
 - g^{\lambda[\nu_1}u_{\mu_1\mu_2}{}^{\nu_2]}
\nn & &
 + \delta_{[\mu_1}{}^{\nu_1}\delta_{\mu_2]}{}^{\nu_2}K^\lambda
 + \delta_{[\mu_1}{}^{[\nu_1}Y_{(3)\mu_2]}{}^{\nu_2]\lambda}
 + Y_{(5)\mu_1\mu_2}{}^{\nu_1\nu_2\lambda},
\\
s_{\mu_1\mu_2}{}^{\nu_1\nu_2} & = &
 c\delta_{[\mu_1}{}^{\nu_1}\delta_{\mu_2]}{}^{\nu_2}
 + \frac{1}{m^2}\delta_{[\mu_1}{}^{[\nu_1}\nabla_{\mu_2]}\nabla^{\nu_2]} c
 + \nabla_{[\mu_1}u_{\mu_2]}{}^{\nu_1\nu_2}
\nn & & 
 + 2\delta_{[\mu_1}{}^{[\nu_1}\nabla_{\mu_2]}K^{\nu_2]}
 - \frac{1}{2}\nabla_{[\mu_1}Y_{(3)\mu_2]}{}^{\nu_1\nu_2}.
\eea
$q_{\mu_1\mu_2}{}^{\nu_1\nu_2}$ depends on $D$:
\begin{alignat}{2}
\bullet~D &\neq 3 & \quad
q_{\mu_1\mu_2}{}^{\nu_1\nu_2} & =
 \Big(c+\frac{2}{D}\nabla_\rho K^\rho\Big)\delta_{[\mu_1}{}^{\nu_1}\delta_{\mu_2]}{}^{\nu_2}
\notag\\ & & & \quad
 + 2\delta_{[\mu_1}{}^{[\nu_1}\nabla_{\mu_2]}K^{\nu_2]}
 - \frac{1}{2}\nabla_{[\mu_1}Y_{(3)\mu_2]}{}^{\nu_1\nu_2}.
\\
\bullet~D &=3 & \quad
q_{\mu_1\mu_2}{}^{\nu_1\nu_2} & =
 \Big(c+2\nabla_\rho K^\rho\Big)\delta_{[\mu_1}{}^{\nu_1}\delta_{\mu_2]}{}^{\nu_2} 
 - 2\delta_{[\mu_1}{}^{[\nu_1}\nabla^{\nu_2]}K_{\mu_2]}.
\end{alignat}

\paragraph{Results for $p=3$}
\
In this case we consider $D\ge 4$, and since full analysis is complicated we only give the general solution to
\eqref{Qo3} and \eqref{qo2}. Because those equations do not contain the mass parameter $m$, 
we do not have to consider massless and massive cases separately. The solution consists of 
$c$, $U_{\mu_1\mu_2}{}^{\nu_1\nu_2\nu_3}$, $V_{\mu_1\mu_2\mu_3}{}^{\nu_1\nu_2}$, $K^\mu$, $Y_{(3)\mu_1\mu_2\mu_3}$,
 $Y_{(5)\mu_1\dots\mu_5}$, and $Y_{(7)\mu_1\dots\mu_7}$.
$c$ is an arbitrary function, 
$U_{\mu_1\mu_2}{}^{\nu_1\nu_2\nu_3}$ and $V_{\mu_1\mu_2\mu_3}{}^{\nu_1\nu_2}$ are arbitrary tensors, and
\begin{alignat}{2}
\bullet~D &=4 & \quad & \text{$K^\mu$ is an arbitrary vector,
 $Y_{(3)\mu_1\mu_2\mu_3}=0$, $Y_{(5)\mu_1\dots\mu_5}=0$, and $Y_{(7)\mu_1\dots\mu_7}=0$.}
\notag\\
\bullet~D &=5 & \quad & \text{$K^\mu$ is a conformal Killing vector,
 $Y_{(3)\mu_1\mu_2\mu_3}$ is a coclosed 3-form, }
\notag\\ & & &
 \text{$Y_{(5)\mu_1\dots\mu_5}=0$, and $Y_{(7)\mu_1\dots\mu_7}=0$.}
\notag\\
\bullet~D &=6 & \quad & \text{$K^\mu$ is a conformal Killing vector,
 $Y_{(3)\mu_1\mu_2\mu_3}$ is a Killing-Yano form,}
\notag\\ & & &
 \text{$Y_{(5)\mu_1\dots\mu_5}$ is a coclosed 5-form, and $Y_{(7)\mu_1\dots\mu_7}=0$.}
\notag\\
\bullet~D &=8 & \quad & \text{$K^\mu$ is a conformal Killing vector, and}
\notag\\ & & &
 \text{$Y_{(3)\mu_1\mu_2\mu_3}$, $Y_{(5)\mu_1\dots\mu_5}$, and $Y_{(7)\mu_1\dots\mu_7}$ are conformal Killing-Yano forms.}
\notag\\
\bullet~D &\neq 4,5,6,8 & \quad & \text{$K^\mu$ is a conformal Killing vector, and}
\notag\\ & & &
 \text{$Y_{(3)\mu_1\mu_2\mu_3}$, $Y_{(5)\mu_1\dots\mu_5}$, and $Y_{(7)\mu_1\dots\mu_7}$ are Killing-Yano forms.}
\notag
\end{alignat}
In either case,
\bea
S_{\mu_1\mu_2\mu_3}{}^{\nu_1\nu_2\nu_3\lambda} & = &
 \delta_{[\mu_1}{}^\lambda U_{\mu_2\mu_3]}{}^{\nu_1\nu_2\nu_3}
 + \delta_{[\mu_1}{}^{\nu_1}\delta_{\mu_2}{}^{\nu_2}\delta_{\mu_3]}{}^{\nu_3} K^\lambda
\nn & &
 + \delta_{[\mu_1}{}^{[\nu_1}\delta_{\mu_2}{}^{\nu_2} Y_{(3)\mu_3]}{}^{\nu_3]\lambda}
 + \delta_{[\mu_1}{}^{[\nu_1} Y_{(5)\mu_2\mu_3]}{}^{\nu_2\nu_3]\lambda}
 + Y_{(7)\mu_1\dots\mu_3}{}^{\nu_1\dots\nu_3\lambda},
\\
Q_{\mu_1\mu_2\mu_3}{}^{\nu_1\nu_2\nu_3\lambda} & = & 
 g^{\lambda[\nu_1}V_{\mu_1\mu_2\mu_3}{}^{\nu_2\nu_3]}
 + \delta_{[\mu_1}{}^{\nu_1}\delta_{\mu_2}{}^{\nu_2}\delta_{\mu_3]}{}^{\nu_3} K^\lambda
\nn & &
 + \delta_{[\mu_1}{}^{[\nu_1}\delta_{\mu_2}{}^{\nu_2} Y_{(3)\mu_3]}{}^{\nu_3]\lambda}
 + \delta_{[\mu_1}{}^{[\nu_1} Y_{(5)\mu_2\mu_3]}{}^{\nu_2\nu_3]\lambda}
 + Y_{(7)\mu_1\dots\mu_3}{}^{\nu_1\dots\nu_3\lambda},
\\
s_{\mu_1\mu_2\mu_3}{}^{\nu_1\nu_2\nu_3} & = &
 c\delta_{[\mu_1}{}^{\nu_1}\delta_{\mu_2}{}^{\nu_2}\delta_{\mu_3]}{}^{\nu_3}
 + \nabla_{[\mu_1}U_{\mu_2\mu_3]}{}^{\nu_1\nu_2\nu_3}
\nn & & 
 + 3\delta_{[\mu_1}{}^{[\nu_1}\delta_{\mu_2}{}^{\nu_2}\nabla_{\mu_3]}K^{\nu_3]}
 - \delta_{[\mu_1}{}^{[\nu_1}\nabla_{\mu_2}Y_{(3)\mu_3]}{}^{\nu_2\nu_3]}
\nn & & 
 + \frac{1}{3}\nabla_{[\mu_1}Y_{(5)\mu_2\mu_3]}{}^{\nu_1\nu_2\nu_3}.
\eea
$q_{\mu_1\mu_2\mu_3}{}^{\nu_1\nu_2\nu_3}$ depends on $D$:
\begin{alignat}{2}
\bullet~D &\neq 4,5,6,8 & \quad
q_{\mu_1\mu_2\mu_3}{}^{\nu_1\nu_2\nu_3} & =
 \Big(c+\frac{2}{D}\nabla_\rho K^\rho\Big)\delta_{[\mu_1}{}^{\nu_1}\delta_{\mu_2}{}^{\nu_2}\delta_{\mu_3]}{}^{\nu_3}
\notag\\ & & & \quad 
 + 3\delta_{[\mu_1}{}^{[\nu_1}\delta_{\mu_2}{}^{\nu_2}\nabla_{\mu_3]}K^{\nu_3]}
 - \delta_{[\mu_1}{}^{[\nu_1}\nabla_{\mu_2}Y_{(3)\mu_3]}{}^{\nu_2\nu_3]}
\notag\\ & & & \quad 
 + \frac{1}{3}\nabla_{[\mu_1}Y_{(5)\mu_2\mu_3]}{}^{\nu_1\nu_2\nu_3}.
\\
\bullet~D &=4 & \quad
q_{\mu_1\mu_2\mu_3}{}^{\nu_1\nu_2\nu_3} & = 
 \Big(c+2\nabla_\rho K^\rho\Big)\delta_{[\mu_1}{}^{\nu_1}\delta_{\mu_2}{}^{\nu_2}\delta_{\mu_3]}{}^{\nu_3} 
 - 3\delta_{[\mu_1}{}^{[\nu_1}\delta_{\mu_2}{}^{\nu_2}\nabla^{\nu_3]}K_{\mu_3]}.
\\
\bullet~D &=5 & \quad
q_{\mu_1\mu_2\mu_3}{}^{\nu_1\nu_2\nu_3} & =
 \Big(c+\frac{2}{5}\nabla_\rho K^\rho\Big)\delta_{[\mu_1}{}^{\nu_1}\delta_{\mu_2}{}^{\nu_2}\delta_{\mu_3]}{}^{\nu_3}
\notag\\ & & & \quad 
 + 3\delta_{[\mu_1}{}^{[\nu_1}\delta_{\mu_2}{}^{\nu_2}\nabla_{\mu_3]}K^{\nu_3]}
 - \delta_{[\mu_1}{}^{[\nu_1}\nabla^{\nu_2}Y_{(3)}{}^{\nu_3]}{}_{\mu_2\mu_3]}.
\\
\bullet~D &=6 & \quad
q_{\mu_1\mu_2\mu_3}{}^{\nu_1\nu_2\nu_3} & =
 \Big(c+\frac{1}{3}\nabla_\rho K^\rho\Big)\delta_{[\mu_1}{}^{\nu_1}\delta_{\mu_2}{}^{\nu_2}\delta_{\mu_3]}{}^{\nu_3}
\notag\\ & & & \quad 
 + 3\delta_{[\mu_1}{}^{[\nu_1}\delta_{\mu_2}{}^{\nu_2}\nabla_{\mu_3]}K^{\nu_3]}
 - \delta_{[\mu_1}{}^{[\nu_1}\nabla_{\mu_2}Y_{(3)\mu_3]}{}^{\nu_2\nu_3]}
\notag\\ & & & \quad 
 - \frac{1}{3}\nabla^{[\nu_1}Y_{(5)}{}^{\nu_2\nu_3]}{}_{\mu_1\mu_2\mu_3}.
\\
\bullet~D &=8 & \quad
q_{\mu_1\mu_2\mu_3}{}^{\nu_1\nu_2\nu_3} & =
 \Big(c+\frac{1}{4}\nabla_\rho K^\rho\Big)\delta_{[\mu_1}{}^{\nu_1}\delta_{\mu_2}{}^{\nu_2}\delta_{\mu_3]}{}^{\nu_3}
\notag\\ & & & \quad 
 + 3\delta_{[\mu_1}{}^{[\nu_1}\delta_{\mu_2}{}^{\nu_2}\nabla_{\mu_3]}K^{\nu_3]}
 - \delta_{[\mu_1}{}^{[\nu_1}\nabla_{\mu_2}Y_{(3)\mu_3]}{}^{\nu_2\nu_3]}
\notag\\ & & & \quad 
 + \frac{1}{3}\nabla_{[\mu_1}Y_{(5)\mu_2\mu_3]}{}^{\nu_1\nu_2\nu_3}
\notag\\ & & & \quad 
 + \frac{1}{3}\delta_{[\mu_1}{}^{[\nu_1}\delta_{\mu_2}{}^{\nu_2}\nabla_{|\rho|}Y_{(3)\mu_3]}{}^{\nu_3]\rho}
 + \frac{1}{2}\delta_{[\mu_1}{}^{[\nu_1}\nabla_{|\rho|}Y_{(5)\mu_2\mu_3]}{}^{\nu_2\nu_3]\rho}
\notag\\ & & & \quad 
 + \nabla_\rho Y_{(7)\mu_1\mu_2\mu_3}{}^{\nu_1\nu_2\nu_3\rho}.
\end{alignat}

\section{A Solution for Arbitrary $p$}

The results of the previous section enable us to give a probable form of solution to the conditions 
\eqref{Qo3}-\eqref{O0} for arbitrary $p$, and we show that it indeed solves the conditions.
Though it is not the general solution, it covers most of the solutions in the previous section.

$S_{\mu_1\dots\mu_p}{}^{\nu_1\dots\nu_p\lambda}$ in the solutions of the previous section suggests
that the following is a solution to \eqref{Qo3} for arbitrary $p$:
\bea
S_{\mu_1\dots\mu_p}{}^{\nu_1\dots\nu_p\lambda} & = &
 \delta_{[\mu_1}{}^\lambda U_{\mu_2\dots\mu_p]}{}^{\nu_1\dots\nu_p}
\nn & &
 + \sum_{n=0}^p\delta_{[\mu_1}{}^{[\nu_1}\delta_{\mu_2}{}^{\nu_2}\dots\delta_{\mu_n}{}^{\nu_n}
 Y_{(2p-2n+1)\mu_{n+1}\dots\mu_p]}{}^{\nu_{n+1}\dots\nu_p]\lambda}.
\label{solS}
\eea
At this stage we only asuume that $Y_{(2n+1)\mu_1\dots\mu_n}{}^{\nu_1\dots\nu_n\lambda}$ are arbitrary 
antisymmetric tensors, and $U_{\mu_1\dots\mu_{p-1}}{}^{\nu_1\dots\nu_p}$ is an arbitrary tensor
except that the first $p-1$ and the next $p$ indices are antisymmetrized.
For $n\ge D/2$, $Y_{(2n+1)\mu_1\dots\mu_n}{}^{\nu_1\dots\nu_n\lambda}$ vanishes.
$Y_{(1)\mu}$ corresponds to $K^\mu$ in the previous section.
If we assume that $S_{\mu_1\dots\mu_p}{}^{\nu_1\dots\nu_p\lambda}$ is in this form,
then from \eqref{Qo3-2},
\bea
Q_{\mu_1\dots\mu_p}{}^{\nu_1\dots\nu_p\lambda} & = & 
 g^{\lambda[\nu_1}V_{\mu_1\dots\mu_p}{}^{\nu_2\dots\nu_p]}
\nn & &
 + \sum_{n=0}^p\delta_{[\mu_1}{}^{[\nu_1}\delta_{\mu_2}{}^{\nu_2}\dots\delta_{\mu_n}{}^{\nu_n}
 Y_{(2p-2n+1)\mu_{n+1}\dots\mu_p]}{}^{\nu_{n+1}\dots\nu_p]\lambda},
\label{solQ}
\eea
where
\bea
V_{\mu_1\dots\mu_p}{}^{\nu_1\dots\nu_{p-1}} & = &
 \frac{p}{D-p+1}Q_{\mu_1\dots\mu_p}{}^{\rho\nu_1\dots\nu_{p-1}}{}_\rho
\nn & &
 -\frac{1}{D-p+1}\sum_{n=0}^{p-1}(n+1)\delta_{[\mu_2}{}^{[\nu_1}\delta_{\mu_3}{}^{\nu_2}\dots\delta_{\mu_{n+1}}{}^{\nu_n}
\nn & & 
 \times Y_{(2p-2n-1)\mu_{n+2}\dots\mu_p}{}^{\nu_{n+1}\dots\nu_{p-1}]}{}_{\mu_1]}.
\eea
A straightforward calculation shows that \eqref{solS} and \eqref{solQ} give a solution to \eqref{Qo3}.
$V_{\mu_1\dots\mu_p}{}^{\nu_1\dots\nu_{p-1}}$ is also an arbitrary tensor
 except that the first $p$ and the next $p-1$ indices are antisymmetrized.

Then we further assume that $Y_{(2n+1)\mu_1\dots\mu_n}{}^{\nu_1\dots\nu_n\lambda}$ are 
Killing-Yano forms for $1\le n\le p$ and $Y_{(1)}{}^\lambda$ is a conformal Killing vector.
From \eqref{qo2-q0},
\beq
q_{\mu_1\dots\mu_p}{}^{\nu_1\dots\nu_p} = 
 s_{\mu_1\dots\mu_p}{}^{\nu_1\dots\nu_p} - \nabla_{[\mu_1}U_{\mu_2\dots\mu_p]}{}^{\nu_1\dots\nu_p}
 + \frac{2}{D}\nabla_\rho Y_{(1)}{}^\rho\delta_{[\mu_1}{}^{\nu_1}\delta_{\mu_2}{}^{\nu_2}\dots\delta_{\mu_p]}{}^{\nu_p},
\label{qo2-q1}
\eeq
and \eqref{qo2-s} uniquely determines $s_{\mu_1\dots\mu_p}{}^{\nu_1\dots\nu_p}$.
However, because the explicit expression of \eqref{qo2-s} for arbitrary $p$ is complicated,
we determine $s_{\mu_1\dots\mu_p}{}^{\nu_1\dots\nu_p}$ in the following indirect way:
from \eqref{qo2-s}, we see that $s_{\mu_1\dots\mu_p}{}^{\nu_1\dots\nu_p}$ is a linear combination of 
the following terms:
\begin{itemize}
\item terms proportional to $\nabla_\gamma U_{\alpha_1\dots\alpha_{p-1}}{}^{\beta_1\dots\beta_p}$
\item $s_{\rho_1\dots\rho_p}{}^{\rho_1\dots\rho_p}\delta_{[\mu_1}{}^{\nu_1}\delta_{\mu_2}{}^{\nu_2}\dots\delta_{\mu_p]}{}^{\nu_p}$
\item $\nabla_\rho Y_{(1)}{}^\rho\delta_{[\mu_1}{}^{\nu_1}\delta_{\mu_2}{}^{\nu_2}\dots\delta_{\mu_p]}{}^{\nu_p}$
\item $\delta_{[\mu_1}{}^{[\nu_1}\delta_{\mu_2}{}^{\nu_2}\dots\delta_{\mu_n}{}^{\nu_n}
 \nabla_{\mu_{n+1}}Y_{(2p-2n-1)\mu_{n+2}\dots\mu_p]}{}^{\nu_{n+1}\dots\nu_p]}$
\end{itemize}
First we determine how $q_{\mu_1\dots\mu_p}{}^{\nu_1\dots\nu_p}$ depends on 
$U_{\alpha_1\dots\alpha_{p-1}}{}^{\beta_1\dots\beta_p}$.
By eliminating $S_{\mu_1\dots\mu_p}{}^{\nu_1\dots\nu_p\lambda}$ and $s_{\mu_1\dots\mu_p}{}^{\nu_1\dots\nu_p}$
from \eqref{qo2} using \eqref{solS} and \eqref{qo2-q1}, 
\bea
\lefteqn{
pq_{[\mu_1\dots\mu_{p-1}}{}^{(\lambda_1|\nu_1\dots\nu_p|}\delta_{\mu_p]}{}^{\lambda_2)}
-pq_{\mu_1\dots\mu_p}{}^{[\nu_1\dots\nu_{p-1}|(\lambda_1}g^{\lambda_2)|\nu_p]}
} \nn & = & 
\text{(terms proportional to $Y_{(2n+1)\alpha_1\dots\alpha_n}{}^{\beta_1\dots\beta_n\gamma}$)}.
\eea
Note that $U_{\alpha_1\dots\alpha_{p-1}}{}^{\beta_1\dots\beta_p}$ does not appear in this equation.
By contracting $\mu_1$ and $\lambda_2$ in this equation,
\bea
& &
(D-p+1)q_{\mu_1\dots\mu_{p-1}}{}^{\lambda\nu_1\dots\nu_p}
-(p+1)q_{\mu_1\dots\mu_{p-1}}{}^{[\lambda\nu_1\dots\nu_p]}
+pq_{\mu_1\dots\mu_{p-1}\rho}{}^{[\nu_1\dots\nu_{p-1}|\rho|}g^{\nu_p]\lambda}
\nn & & 
=\text{(terms proportional to $Y_{(2n+1)\alpha_1\dots\alpha_n}{}^{\beta_1\dots\beta_n\gamma}$)}.
\eea
By making the same manipulation as \eqref{qo2-aux1}-\eqref{qo2-s} to the above equation,
we obtain
\bea
q_{\mu_1\dots\mu_p}{}^{\nu_1\dots\nu_p} & = & 
\frac{p!(D-p)!}{D!}q_{\rho_1\dots\rho_p}{}^{\rho_1\dots\rho_p}
 \delta_{[\mu_1}{}^{\nu_1}\dots\delta_{\mu_p]}{}^{\nu_p}
\nn & & 
+\text{(terms proportional to $Y_{(2n+1)\alpha_1\dots\alpha_n}{}^{\beta_1\dots\beta_n\gamma}$)},
\eea
and therefore in the right hand side of the above $U_{\alpha_1\dots\alpha_{p-1}}{}^{\beta_1\dots\beta_p}$
can be present only in the first term.
Then $s_{\mu_1\dots\mu_p}{}^{\nu_1\dots\nu_p}$ and $q_{\mu_1\dots\mu_p}{}^{\nu_1\dots\nu_p}$ 
can be put in the following form:
\bea
s_{\mu_1\dots\mu_p}{}^{\nu_1\dots\nu_p} & = &
 c\delta_{[\mu_1}{}^{\nu_1}\delta_{\mu_2}{}^{\nu_2}\dots\delta_{\mu_p]}{}^{\nu_p}
 + \nabla_{[\mu_1}U_{\mu_2\dots\mu_p]}{}^{\nu_1\dots\nu_p}
\nn & & 
 + \sum_{n=0}^{p-1}e_n\delta_{[\mu_1}{}^{[\nu_1}\delta_{\mu_2}{}^{\nu_2}\dots\delta_{\mu_n}{}^{\nu_n}
 \nabla_{\mu_{n+1}}Y_{(2p-2n-1)\mu_{n+2}\dots\mu_p]}{}^{\nu_{n+1}\dots\nu_p]},
\\
q_{\mu_1\dots\mu_p}{}^{\nu_1\dots\nu_p} & = &
 \Big(c+\frac{2}{D}\nabla_\rho Y_{(1)}{}^\rho\Big)\delta_{[\mu_1}{}^{\nu_1}\delta_{\mu_2}{}^{\nu_2}\dots\delta_{\mu_p]}{}^{\nu_p}
\nn & & 
 + \sum_{n=0}^{p-1}e_n\delta_{[\mu_1}{}^{[\nu_1}\delta_{\mu_2}{}^{\nu_2}\dots\delta_{\mu_n}{}^{\nu_n}
 \nabla_{\mu_{n+1}}Y_{(2p-2n-1)\mu_{n+2}\dots\mu_p]}{}^{\nu_{n+1}\dots\nu_p]},
\eea
where $c$ is a scalar function, and $e_n$ are constants.
From \eqref{qo2}, by relatively easier calculation it turns out that
$e_n=(-1)^{p+n+1}\frac{n+1}{p-n}$, and
\bea
s_{\mu_1\dots\mu_p}{}^{\nu_1\dots\nu_p} & = &
 c\delta_{[\mu_1}{}^{\nu_1}\delta_{\mu_2}{}^{\nu_2}\dots\delta_{\mu_p]}{}^{\nu_p}
 + \nabla_{[\mu_1}U_{\mu_2\dots\mu_p]}{}^{\nu_1\dots\nu_p}
\nn & & 
 + \sum_{n=0}^{p-1}(-1)^{p+n+1}\frac{n+1}{p-n}\delta_{[\mu_1}{}^{[\nu_1}\delta_{\mu_2}{}^{\nu_2}\dots\delta_{\mu_n}{}^{\nu_n}
\nn & & 
 \times\nabla_{\mu_{n+1}}Y_{(2p-2n-1)\mu_{n+2}\dots\mu_p]}{}^{\nu_{n+1}\dots\nu_p]},
\\
q_{\mu_1\dots\mu_p}{}^{\nu_1\dots\nu_p} & = &
 \Big(c+\frac{2}{D}\nabla_\rho Y_{(1)}{}^\rho\Big)\delta_{[\mu_1}{}^{\nu_1}\delta_{\mu_2}{}^{\nu_2}\dots\delta_{\mu_p]}{}^{\nu_p}
\nn & & 
 + \sum_{n=0}^{p-1}(-1)^{p+n+1}\frac{n+1}{p-n}\delta_{[\mu_1}{}^{[\nu_1}\delta_{\mu_2}{}^{\nu_2}\dots\delta_{\mu_n}{}^{\nu_n}
\nn & & 
 \times\nabla_{\mu_{n+1}}Y_{(2p-2n-1)\mu_{n+2}\dots\mu_p]}{}^{\nu_{n+1}\dots\nu_p]}.
\eea
These give the solution to \eqref{qo2}.
Next we analyze \eqref{O1}.
We compute the right hand side minus the left hand side of \eqref{O1} and see if it vanishes or gives extra conditions.
First we see only the part dependent on $U_{\alpha_1\dots\alpha_{p-1}}{}^{\beta_1\dots\beta_p}$.
The result of the straightforward calculation is
\begin{align}
& -m^2\delta_{[\mu_1}{}^\lambda U_{\mu_2\dots\mu_p]}{}^{\nu_1\dots\nu_p}
\notag\\
& +pR^\lambda{}_{[\mu_1}{}^{\rho[\nu_1}U_{\mu_2\dots\mu_p]\rho}{}^{\nu_2\dots\nu_p]}
+\frac{1}{2}p(p-1)R_{[\mu_1\mu_2}{}^{\rho[\nu_1}U^{|\lambda|}{}_{\mu_3\dots\mu_p]\rho}{}^{\nu_2\dots\nu_p]}
\notag\\
& +(\nabla^\lambda\nabla_{[\mu_1}-\nabla_{[\mu_1}\nabla^\lambda)U_{\mu_2\dots\mu_p]}{}^{\nu_1\dots\nu_p}
+(p-1)\nabla_{[\mu_1}\nabla_{\mu_2}U^{\lambda}{}_{\mu_3\dots\mu_p]}{}^{\nu_1\dots\nu_p}.
\end{align}
By replacing the commutators of covariant derivatives with the Riemann tensors and using the Bianchi identity,
the above expression is reduced to
\beq
-m^2\delta_{[\mu_1}{}^\lambda U_{\mu_2\dots\mu_p]}{}^{\nu_1\dots\nu_p}.
\eeq
Next we see only the part dependent on $V_{\alpha_1\dots\alpha_p}{}^{\beta_1\dots\beta_{p-1}}$ and $c$.
The result of the straightforward calculation is
\begin{align}
& m^2g^{\lambda[\nu_1}V_{\mu_1\dots\mu_p}{}^{\nu_2\dots\nu_p]}
+2\delta_{[\mu_1}{}^{\nu_1}\delta_{\mu_2}{}^{\nu_2}\dots\delta_{\mu_p]}{}^{\nu_p}\nabla^\lambda c
\notag\\
& -p\delta_{[\mu_1}{}^\lambda\delta_{\mu_2}{}^{[\nu_2}\dots\delta_{\mu_p]}{}^{\nu_p}\nabla^{\nu_1]}c
-pg^{\lambda[\nu_1}\delta_{[\mu_2}{}^{\nu_2}\dots\delta_{\mu_p}{}^{\nu_p]}\nabla_{\mu_1]}c.
\end{align}
Finally we see only the part dependent on $Y_{(2n+1)\alpha_1\dots\alpha_n}{}^{\beta_1\dots\beta_n\gamma}$.
Some terms need attention:
The contribution from the term $-p\nabla^\rho\nabla_{[\mu_1}S_{|\rho|\mu_2\dots\mu_p]}{}^{\nu_1\dots\nu_p\lambda}$ is
\bea
\lefteqn{
-\sum_{n=0}^{p-1}(n+1)\delta_{[\mu_1}{}^{[\nu_1}\delta_{\mu_2}{}^{\nu_2}\dots\delta_{\mu_n}{}^{\nu_n}
  \nabla^{\nu_{n+1}}\nabla_{\mu_{n+1}}Y_{(2p-2n-1)\mu_{n+2}\dots\mu_p]}{}^{\nu_{n+2}\dots\nu_p]\lambda}
} \nn 
\lefteqn{
+\sum_{n=0}^{p-1}(-1)^{p+n}(p-n)\delta_{[\mu_1}{}^{[\nu_1}\delta_{\mu_2}{}^{\nu_2}\dots\delta_{\mu_n}{}^{\nu_n}
  \nabla^{|\rho|}\nabla_{\mu_{n+1}}Y_{(2p-2n+1)\mu_{n+2}\dots\mu_p]\rho}{}^{\nu_{n+1}\dots\nu_p]\lambda}
} \nn & = &
 \sum_{n=0}^{p-2}(-1)^{p+n+1}(n+1)\delta_{[\mu_1}{}^{[\nu_1}\delta_{\mu_2}{}^{\nu_2}\dots\delta_{\mu_n}{}^{\nu_n}
  \nabla^{\nu_{n+1}}\nabla^{\nu_{n+2}}Y_{(2p-2n-1)\mu_{n+1}\dots\mu_p]}{}^{\nu_{n+3}\dots\nu_p]\lambda}
\nn & &
+\sum_{n=0}^{p-1}(p-n)\delta_{[\mu_1}{}^{[\nu_1}\delta_{\mu_2}{}^{\nu_2}\dots\delta_{\mu_n}{}^{\nu_n}
  \nabla^2Y_{(2p-2n+1)\mu_{n+1}\dots\mu_p]}{}^{\nu_{n+1}\dots\nu_p]\lambda}
\nn & &
 -p\delta_{[\mu_1}{}^{[\nu_1}\delta_{\mu_2}{}^{\nu_2}\dots\delta_{\mu_{p-1}}{}^{\nu_{p-1}}
  \nabla^{\nu_p]}\nabla_{\mu_p]}Y_{(1)}{}^\lambda.
\label{O1-aux1}
\eea
The contribution from the term $-p\nabla^\rho s_{\rho[\mu_2\dots\mu_p}{}^{\nu_1\dots\nu_p}\delta_{\mu_1]}{}^\lambda$ is
\bea
\lefteqn{
\sum_{n=0}^{p-2}\frac{(n+1)(n+2)}{p-n-1}(-1)^{p+1}
 \delta_{[\mu_1}{}^\lambda\delta_{\mu_2}{}^{[\nu_1}\dots\delta_{\mu_{n+1}}{}^{\nu_n}
 \nabla^{\nu_{n+1}}\nabla_{\mu_{n+2}}Y_{(2p-2n-3)\mu_{n+3}\dots\mu_p]}{}^{\nu_{n+2}\dots\nu_p]}
} \nn
\lefteqn{ +\sum_{n=0}^{p-2}\frac{(n+1)(p-n-1)}{p-n}(-1)^{n+1}
  \delta_{[\mu_1}{}^\lambda\delta_{\mu_2}{}^{[\nu_1}\dots\delta_{\mu_{n+1}}{}^{\nu_n}
  \nabla^{|\rho|}\nabla_{\mu_{n+2}}Y_{(2p-2n-1)\mu_{n+3}\dots\mu_p]\rho}{}^{\nu_{n+1}\dots\nu_p]}
}\nn
\lefteqn{ +\sum_{n=0}^{p-1}\frac{n+1}{p-n}(-1)^p
  \delta_{[\mu_1}{}^\lambda\delta_{\mu_2}{}^{[\nu_1}\dots\delta_{\mu_{n+1}}{}^{\nu_n}
  \nabla^2Y_{(2p-2n-1)\mu_{n+2}\dots\mu_p]\rho}{}^{\nu_{n+1}\dots\nu_p]}
} \nn & = &
\sum_{n=0}^{p-2}\frac{(n+1)(n+2)}{p-n-1}(-1)^n
 \delta_{[\mu_1}{}^\lambda\delta_{\mu_2}{}^{[\nu_1}\dots\delta_{\mu_{n+1}}{}^{\nu_n}
 \nabla^{\nu_{n+1}}\nabla^{\nu_{n+2}}Y_{(2p-2n-3)\mu_{n+2}\dots\mu_p]}{}^{\nu_{n+3}\dots\nu_p]}
\nn & &
 +\sum_{n=0}^{p-1}(n+1)(-1)^p
  \delta_{[\mu_1}{}^\lambda\delta_{\mu_2}{}^{[\nu_1}\dots\delta_{\mu_{n+1}}{}^{\nu_n}
  \nabla^2Y_{(2p-2n-1)\mu_{n+2}\dots\mu_p]}{}^{\nu_{n+1}\dots\nu_p]}
\nn & &
 -\frac{2}{D}p(p-1)\delta_{[\mu_1}{}^\lambda\delta_{\mu_2}{}^{[\nu_2}\dots\delta_{\mu_p]}{}^{\nu_p}
  \nabla^{\nu_1]}\nabla_\rho Y_{(1)}{}^\rho.
\label{O1-aux2}
\eea
The contribution from the term $-p\nabla_{[\mu_1}s^\lambda{}_{\mu_2\dots\mu_p]}{}^{\nu_1\dots\nu_p}$ is
\bea
\lefteqn{
\sum_{n=0}^{p-2}\frac{(n+1)(n+2)}{p-n-1}(-1)^{p+1}
 g^{\lambda[\nu_1}\delta_{[\mu_1}{}^{\nu_2}\dots\delta_{\mu_n}{}^{\nu_{n+1}}
 \nabla_{\mu_{n+1}}\nabla_{\mu_{n+2}}Y_{(2p-2n-3)\mu_{n+3}\dots\mu_p]}{}^{\nu_{n+2}\dots\nu_p]}
} \nn
\lefteqn{
-\sum_{n=0}^{p-2}\frac{(n+1)(p-n-1)}{p-n}
 \delta_{[\mu_1}{}^{[\nu_1}\dots\delta_{\mu_n}{}^{\nu_n}
 \nabla_{\mu_{n+1}}\nabla_{\mu_{n+2}}Y_{(2p-2n-1)\mu_{n+3}\dots\mu_p]}{}^{|\lambda|\nu_{n+1}\dots\nu_p]}
} \nn
\lefteqn{
+\sum_{n=0}^{p-1}\frac{n+1}{p-n}(-1)^{p+n}
 \delta_{[\mu_1}{}^{[\nu_1}\dots\delta_{\mu_n}{}^{\nu_n}
 \nabla_{\mu_{n+1}}\nabla^{|\lambda|}Y_{(2p-2n-1)\mu_{n+2}\dots\mu_p]}{}^{\nu_{n+1}\dots\nu_p]}
} \nn & = & 
\sum_{n=0}^{p-2}\frac{(n+1)(n+2)}{p-n-1}(-1)^{p+1}
 g^{\lambda[\nu_1}\delta_{[\mu_1}{}^{\nu_2}\dots\delta_{\mu_n}{}^{\nu_{n+1}}
 \nabla_{\mu_{n+1}}\nabla_{\mu_{n+2}}Y_{(2p-2n-3)\mu_{n+3}\dots\mu_p]}{}^{\nu_{n+2}\dots\nu_p]}
\nn & &
-\sum_{n=0}^{p-2}(n+1)
 \delta_{[\mu_1}{}^{[\nu_1}\dots\delta_{\mu_n}{}^{\nu_n}
 \nabla_{\mu_{n+1}}\nabla_{\mu_{n+2}}Y_{(2p-2n-1)\mu_{n+3}\dots\mu_p]}{}^{|\lambda|\nu_{n+1}\dots\nu_p]}
\nn & &
-p\delta_{[\mu_1}{}^{[\nu_1}\dots\delta_{\mu_{p-1}}{}^{\nu_{p-1}}
 \nabla_{\mu_p]}\nabla^{|\lambda|}Y_{(1)}{}^{\nu_p]}.
\label{O1-aux3}
\eea
In \eqref{O1-aux1}, \eqref{O1-aux2}, and \eqref{O1-aux3},
the first expressions are calculated without using the assumption that 
$Y_{(2n+1)\mu_1\dots\mu_n}{}^{\nu_1\dots\nu_n\lambda}$ are 
Killing-Yano forms for $1\le n\le p$ and $Y_{(1)}{}^\lambda$ is a conformal Killing vector.
The second expressions are given by using the fact that for $n\ge 1$
any two indices in $\nabla_\delta Y_{(2n+1)\alpha_1\dots\alpha_n}{}^{\beta_1\dots\beta_n\gamma}$ can be freely exchanged
with extra minus sign, and for $n=0$ we have an additional term proportional to $\nabla_\rho Y_{(1)}{}^\rho$ upon the exchange.
Then those expressions are simplified further by replacing the commutators of covariant derivatives
with the Riemann tensors and by
\beq
\nabla^2 Y_{(2n+1)\mu_1\dots\mu_{2n+1}} = -R_{\rho[\mu_1}Y_{(2n+1)}{}^\rho{}_{\mu_2\dots\mu_{2n+1}]}
 + nR_{\rho\sigma[\mu_1\mu_2}Y_{(2n+1)}{}^{\rho\sigma}{}_{\mu_3\dots\mu_{2n+1}]},
\label{ddKYO1}
\eeq
\beq
\nabla_\mu\nabla_\nu Y_{(1)\lambda} = R_{\lambda\nu\mu\rho}Y_{(1)}{}^\rho
 + \frac{1}{D}( g_{\lambda\mu}\nabla_\nu\nabla_\rho Y_{(1)}{}^\rho
  + g_{\lambda\nu}\nabla_\mu\nabla_\rho Y_{(1)}{}^\rho
  - g_{\mu\nu}\nabla_\lambda\nabla_\rho Y_{(1)}{}^\rho ).
\eeq
(See \eqref{ddCKY} in Appendix.)
The contribution from the term $2\nabla^\lambda s_{\mu_1\dots\mu_p}{}^{\nu_1\dots\nu_p}$ is
simplified by
\beq
\nabla_\nu\nabla_\lambda Y_{(2n+1)\mu_1\dots\mu_{2n+1}} = 
 -(n+1)R_{[\lambda\mu_1|\nu}{}^\rho Y_{(2n+1)\rho|\mu_2\dots\mu_{2n+1}]}.
\eeq
The contribution from the term $\nabla^2S_{\mu_1\dots\mu_p}{}^{\nu_1\dots\nu_p\lambda}$ is
simplified by \eqref{ddKYO1} and
\beq
\nabla^2 Y_{(1)\mu} = -R_{\rho\mu}Y_{(1)}{}^\rho - \frac{D-2}{D}\nabla_\mu\nabla_\rho Y_{(1)}{}^\rho.
\label{ddCKYO1}
\eeq
Other terms can be calculated straightforwardly. The total result for the part dependent on
$Y_{(2n+1)\alpha_1\dots\alpha_n}{}^{\beta_1\dots\beta_n\gamma}$ is 
\begin{align}
& -\sum_{n=0}^{p-2}2(n+1)
 \delta_{[\mu_1}{}^{[\nu_1}\dots\delta_{\mu_n}{}^{\nu_n}
 I_{(2p-2n-1)\mu_{n+1}}{}^{\nu_{n+1}}{}^{|\lambda|}{}_{\mu_{n+2}\dots\mu_p]}{}^{\nu_{n+2}\dots\nu_p]}
\notag\\
& +\sum_{n=0}^{p-1}(-1)^{p+1}\frac{p-n}{2p-2n+1}
 \delta_{[\mu_1}{}^{[\nu_2}\dots\delta_{\mu_n}{}^{\nu_{n+1}}
\notag\\ 
& \times\Big[
 J_{(2p-2n+1)}{}^{|\lambda|}{}_{\mu_{n+1}}{}^{\nu_1}{}_{\mu_{n+2}\dots\mu_p]}{}^{\nu_{n+2}\dots\nu_p]}
 -(p-n)J_{(2p-2n+1)}{}^{\nu_1}{}_{\mu_{n+1}}{}^{|\lambda|}{}_{\mu_{n+2}\dots\mu_p]}{}^{\nu_{n+2}\dots\nu_p]}
\Big]
\notag\\
& +\sum_{n=0}^{p-2}(-1)^{p+1}\frac{(n+1)(p-n-1)}{2p-2n-1}
 \delta_{[\mu_1}{}^\lambda\delta_{\mu_2}{}^{[\nu_1}\dots\delta_{\mu_{n+1}}{}^{\nu_n}
 J_{(2p-2n-1)\mu_{n+2}}{}^{\nu_{n+1}}{}^{\nu_{n+2}}{}_{\mu_{n+3}\dots\mu_p]}{}^{\nu_{n+3}\dots\nu_p]}
\notag\\
& +\frac{D-2p-2}{D}(p\delta_{[\mu_1}{}^\lambda\delta_{\mu_2}{}^{[\nu_2}\dots\delta_{\mu_p]}{}^{\nu_p}\nabla^{\nu_1]}
 -\delta_{[\mu_1}{}^{\nu_1}\dots\delta_{\mu_p]}{}^{\nu_p}\nabla^\lambda
)\nabla_\rho Y_{(1)}{}^\rho,
\end{align}
where
\bea
I_{(2n+1)\nu_1\nu_2\mu_1\dots\mu_{2n+1}} & = & R_{\nu_1\nu_2[\mu_1}{}^\rho Y_{(2n+1)|\rho|\mu_2\dots\mu_{2n+1}]}
\nn & &
 + \frac{1}{2}R_{[\mu_1\mu_2|\nu_1}{}^\rho Y_{(2n+1)\rho\nu_2|\mu_3\dots\mu_{2n+1}]}
\nn & &
 - \frac{1}{2}R_{[\mu_1\mu_2|\nu_2}{}^\rho Y_{(2n+1)\rho\nu_1|\mu_3\dots\mu_{2n+1}]},
\\
J_{(2n+1)\nu_1\nu_2\mu_1\dots\mu_{2n+1}} & = & R_{\nu_1\rho}Y_{(2n+1)}{}^\rho{}_{\nu_2\mu_1\dots\mu_{2n-1}}
\nn & &
 + \frac{1}{2}(2n-1)R_{\rho_1\rho_2\nu_1[\mu_1}Y_{(2n+1)}{}^{\rho_1\rho_2}{}_{|\nu_2|\mu_2\dots\mu_{2n-1}]}
 + (\nu_1\leftrightarrow\nu_2).
\eea
These vanish because of \eqref{CKYint1} and \eqref{CKYint2}.
Then \eqref{O1} is reduced to
\bea
0 & = & m^2(g^{\lambda[\nu_1}V_{\mu_1\dots\mu_p}{}^{\nu_2\dots\nu_p]}
 -\delta_{[\mu_1}{}^\lambda U_{\mu_2\dots\mu_p]}{}^{\nu_1\dots\nu_p})
\nn & &
-pg^{\lambda[\nu_1}\delta_{[\mu_2}{}^{\nu_2}\dots\delta_{\mu_p}{}^{\nu_p]}\nabla_{\mu_1]}c
-p\delta_{[\mu_1}{}^\lambda\delta_{\mu_2}{}^{[\nu_2}\dots\delta_{\mu_p]}{}^{\nu_p}\nabla^{\nu_1]}
 \Big(c-\frac{D-2p-2}{D}\nabla_\rho Y_{(1)}{}^\rho\Big)
\nn & &
+\delta_{[\mu_1}{}^{\nu_1}\dots\delta_{\mu_p]}{}^{\nu_p}\nabla^\lambda
 \Big(2c-\frac{D-2p-2}{D}\nabla_\rho Y_{(1)}{}^\rho\Big).
\label{O1-1}
\eea
By contracting $(\mu_1,\dots,\mu_p)$ and $(\nu_1,\dots,\nu_p)$ in \eqref{O1-1},
\bea
0 & = & m^2(V^\lambda{}_{\rho_1\dots\rho_{p-1}}{}^{\rho_1\dots\rho_{p-1}}
 -U_{\rho_1\dots\rho_{p-1}}{}^{\lambda\rho_1\dots\rho_{p-1}})
\nn & &
+\frac{(D-1)!}{p!(D-p-1)!}\nabla^\lambda\Big(2c-\frac{D-2p-2}{D}\nabla_\rho Y_{(1)}{}^\rho\Big).
\eea
By contracting $(\lambda,\mu_2,\dots,\mu_p)$ and $(\nu_1,\dots,\nu_p)$ in \eqref{O1-1},
\bea
0 & = & m^2[(D-p+1)V^\lambda{}_{\rho_1\dots\rho_{p-1}}{}^{\rho_1\dots\rho_{p-1}}
 -U_{\rho_1\dots\rho_{p-1}}{}^{\lambda\rho_1\dots\rho_{p-1}}]
\nn & &
-\frac{(D-1)!}{(p-1)!(D-p-1)!}\nabla^\lambda c.
\eea
By contracting $(\mu_1,\dots,\mu_p)$ and $(\lambda,\nu_2,\dots,\nu_p)$ in \eqref{O1-1},
\bea
0 & = & m^2[V^\lambda{}_{\rho_1\dots\rho_{p-1}}{}^{\rho_1\dots\rho_{p-1}}
 -(D-p+1)U_{\rho_1\dots\rho_{p-1}}{}^{\lambda\rho_1\dots\rho_{p-1}}]
\nn & &
-\frac{(D-1)!}{(p-1)!(D-p-1)!}\nabla^\lambda\Big(c-\frac{D-2p-2}{D}\nabla_\rho Y_{(1)}{}^\rho\Big).
\eea
From these three equations,
\bea
0 & = & \nabla^\lambda\Big(2c-\frac{D-2p-2}{D}\nabla_\rho Y_{(1)}{}^\rho\Big),
\label{O1-c}
\\
m^2V^\lambda{}_{\rho_1\dots\rho_{p-1}}{}^{\rho_1\dots\rho_{p-1}} & = & 
 \frac{(D-1)!}{(p-1)!(D-p)!}\nabla^\lambda c,
\label{O1-aux4}
\\
m^2U_{\rho_1\dots\rho_{p-1}}{}^{\lambda\rho_1\dots\rho_{p-1}} & = & 
 \frac{(D-1)!}{(p-1)!(D-p)!}\nabla^\lambda c.
\label{O1-aux5}
\eea
If $m=0$, then $c$ is a constant, and $\nabla_\rho Y_{(1)}{}^\rho$ is a constant or $D=2p+2$.
Then the right hand side of \eqref{O1-1} vanishes and \eqref{O1} is solved.
For $m\neq 0$, \eqref{O1-1} is reduced to
\beq
0=g^{\lambda[\nu_1}\wt{V}_{\mu_1\dots\mu_p}{}^{\nu_2\dots\nu_p]}
 -\delta_{[\mu_1}{}^\lambda\wt{U}_{\mu_2\dots\mu_p]}{}^{\nu_1\dots\nu_p},
\label{O1-2}
\eeq
where
\bea
\wt{V}_{\mu_1\dots\mu_p}{}^{\nu_1\dots\nu_{p-1}} & = & V_{\mu_1\dots\mu_p}{}^{\nu_1\dots\nu_{p-1}}
 -\frac{p}{m^2}\delta_{[\mu_2}{}^{\nu_1}\dots\delta_{\mu_p}{}^{\nu_{p-1}}\nabla_{\mu_1]}c,
\\
\wt{U}_{\mu_1\dots\mu_{p-1}}{}^{\nu_1\dots\nu_p} & = & U_{\mu_1\dots\mu_{p-1}}{}^{\nu_1\dots\nu_p}
 -\frac{p}{m^2}\delta_{\mu_1}{}^{[\nu_2}\dots\delta_{\mu_{p-1}}{}^{\nu_p}\nabla^{\nu_1]}c.
\eea
From \eqref{O1-aux4} and \eqref{O1-aux5},
\bea
\wt{V}^\lambda{}_{\rho_1\dots\rho_{p-1}}{}^{\rho_1\dots\rho_{p-1}} & = & 0,
\label{UV0-1}
\\
\wt{U}_{\rho_1\dots\rho_{p-1}}{}^{\lambda\rho_1\dots\rho_{p-1}} & = & 0.
\label{UV0-2}
\eea
By contracting $(\mu_{p-n+1},\dots,\mu_p)$ and $(\nu_{p-n+1},\dots,\nu_p)$ in \eqref{O1-2},
\bea
0 & = &
\wt{V}^\lambda{}_{\mu_1\dots\mu_{p-n}\rho_1\dots\rho_{n-1}}{}^{\nu_1\dots\nu_{p-n}\rho_1\dots\rho_{n-1}}
 -\wt{U}_{\mu_1\dots\mu_{p-n}\rho_1\dots\rho_{n-1}}{}^{\lambda\nu_1\dots\nu_{p-n}\rho_1\dots\rho_{n-1}}
\nn & &
 +\frac{p-n}{n}(
 g^{\lambda[\nu_1}\wt{V}_{\mu_1\dots\mu_{p-n}\rho_1\dots\rho_n}{}^{\nu_2\dots\nu_{p-n}]\rho_1\dots\rho_n}
 -\delta_{[\mu_1}{}^\lambda\wt{U}_{\mu_2\dots\mu_{p-n}]\rho_1\dots\rho_n}{}^{\nu_1\dots\nu_{p-n}\rho_1\dots\rho_n}
).
\label{VU1}
\eea
By contracting $(\lambda,\mu_{p-n+1},\dots,\mu_p)$ and $(\nu_1,\nu_{p-n+1},\dots,\nu_p)$ in \eqref{O1-2},
\bea
0 & = & (D-p+1)\wt{V}_{\mu_1\dots\mu_{p-n}\rho_1\dots\rho_n}{}^{\nu_2\dots\nu_{p-n}\rho_1\dots\rho_n}
\nn & &
 -(p-n)\wt{U}_{[\mu_2\dots\mu_{p-n}|\rho_1\dots\rho_n|\mu_1]}{}^{\nu_2\dots\nu_{p-n}\rho_1\dots\rho_n}.
\label{VU2}
\eea
By contracting $(\mu_1,\mu_{p-n+1},\dots,\mu_p)$ and $(\lambda,\nu_{p-n+1},\dots,\nu_p)$ in \eqref{O1-2},
\bea
0 & = &(D-p+1)\wt{U}_{\mu_2\dots\mu_{p-n}\rho_1\dots\rho_n}{}^{\nu_1\dots\nu_{p-n}\rho_1\dots\rho_n}
\nn & &
 -(p-n)\wt{V}^{[\nu_1}{}_{\mu_2\dots\mu_{p-n}\rho_1\dots\rho_n}{}^{\nu_2\dots\nu_{p-n}]\rho_1\dots\rho_n}.
\label{VU3}
\eea
By replacing $n$ with $n+1$, renaming $\lambda$, $\mu_1,\dots,\mu_{p-n-1}$ to $\mu_1,\dots,\mu_{p-n}$,
antisymmetrizing $\mu_1,\dots,\mu_{p-n}$ in \eqref{VU1}, and using \eqref{VU2},
\bea
\lefteqn{
(D-2p+n+1)\wt{V}_{\mu_1\dots\mu_{p-n}\rho_1\dots\rho_n}{}^{\nu_2\dots\nu_{p-n}\rho_1\dots\rho_n}
} \nn & = &
 \frac{(p-n)(p-n-1)}{n+1}\delta_{[\mu_1}{}^{[\nu_2}
 \wt{V}_{\mu_2\dots\mu_{p-n}]\rho_1\dots\rho_{n+1}}{}^{\nu_3\dots\nu_{p-n}]\rho_1\dots\rho_{n+1}}.
\label{VU2-1}
\eea
By replacing $n$ with $n+1$, renaming $\lambda$, $\nu_1,\dots,\nu_{p-n-1}$ to $\nu_1,\dots,\nu_{p-n}$,
antisymmetrizing $\nu_1,\dots,\nu_{p-n}$ in \eqref{VU1}, and using \eqref{VU3},
\bea
\lefteqn{
(D-2p+n+1)\wt{U}_{\mu_2\dots\mu_{p-n}\rho_1\dots\rho_n}{}^{\nu_1\dots\nu_{p-n}\rho_1\dots\rho_n}
} \nn & = &
 \frac{(p-n)(p-n-1)}{n+1}\delta_{[\mu_2}{}^{[\nu_1}
 \wt{U}_{\mu_3\dots\mu_{p-n}]\rho_1\dots\rho_{n+1}}{}^{\nu_2\dots\nu_{p-n}]\rho_1\dots\rho_{n+1}}.
\label{VU3-1}
\eea
Unless $D-2p+n+1=0$, by using the above two equations we can uniquely determine 
$\wt{V}_{\mu_1\dots\mu_{p-n}\rho_1\dots\rho_n}{}^{\nu_2\dots\nu_{p-n}\rho_1\dots\rho_n}$ and
$\wt{U}_{\mu_2\dots\mu_{p-n}\rho_1\dots\rho_n}{}^{\nu_1\dots\nu_{p-n}\rho_1\dots\rho_n}$
from $\wt{V}_{\mu_2\dots\mu_{p-n}\rho_1\dots\rho_{n+1}}{}^{\nu_3\dots\nu_{p-n}\rho_1\dots\rho_{n+1}}$
and $\wt{U}_{\mu_3\dots\mu_{p-n}\rho_1\dots\rho_{n+1}}{}^{\nu_2\dots\nu_{p-n}\rho_1\dots\rho_{n+1}}$.
Therefore if $2p\le D$, from \eqref{UV0-1} and \eqref{UV0-2}, we can show
$\wt{V}_{\mu_1\dots\mu_p}{}^{\nu_1\dots\nu_{p-1}}=0$ and
$\wt{U}_{\mu_1\dots\mu_{p-1}}{}^{\nu_1\dots\nu_p}=0$,
i.e.
\bea
V_{\mu_1\dots\mu_p}{}^{\nu_1\dots\nu_{p-1}} & = &
 \frac{p}{m^2}\delta_{[\mu_2}{}^{\nu_1}\dots\delta_{\mu_p}{}^{\nu_{p-1}}\nabla_{\mu_1]}c,
\label{VUav}
\\
U_{\mu_1\dots\mu_{p-1}}{}^{\nu_1\dots\nu_p} & = &
 \frac{p}{m^2}\delta_{\mu_1}{}^{[\nu_2}\dots\delta_{\mu_{p-1}}{}^{\nu_p}\nabla^{\nu_1]}c.
\label{VUau}
\eea
If $2p-D-1\ge 0$, 
$\wt{V}_{\mu_1\dots\mu_{D-p+1}\rho_1\dots\rho_{2p-D-1}}{}^{\nu_2\dots\nu_{D-p+1}\rho_1\dots\rho_{2p-D-1}}$
and \\
$\wt{U}_{\mu_2\dots\mu_{D-p+1}\rho_1\dots\rho_{2p-D-1}}{}^{\nu_1\dots\nu_{D-p+1}\rho_1\dots\rho_{2p-D-1}}$
are left undetermined, and for $n\le 2p-D-1$, we can show
$\wt{V}_{\mu_1\dots\mu_{p-n}\rho_1\dots\rho_n}{}^{\nu_2\dots\nu_{p-n}\rho_1\dots\rho_n}=0$ and
$\wt{U}_{\mu_2\dots\mu_{p-n}\rho_1\dots\rho_n}{}^{\nu_1\dots\nu_{p-n}\rho_1\dots\rho_n}=0$.
From \eqref{VU1} for $n=2p-D$,
\bea
\wt{V}^\lambda{}_{\mu_1\dots\mu_{D-p}\rho_1\dots\rho_{2p-D-1}}{}^{\nu_1\dots\nu_{D-p}\rho_1\dots\rho_{2p-D-1}}
& = & \wt{U}_{\mu_1\dots\mu_{D-p}\rho_1\dots\rho_{2p-D-1}}{}^{\lambda\nu_1\dots\nu_{D-p}\rho_1\dots\rho_{2p-D-1}}
\nn & \equiv & (-1)^{D+1}\frac{(D-p)!(D-p+1)!((2p-D-1)!)^2}{p!(p-1)!}
\nn & & \times
 u_{\mu_1\dots\mu_{D-p}}{}^{\lambda\nu_1\dots\nu_{D-p}},
\eea
where $u_{\mu_1\dots\mu_{2D-2p+1}}$ is an antisymmetric tensor. Then
\bea
V_{\mu_1\dots\mu_p}{}^{\nu_1\dots\nu_{p-1}} & = &
 \frac{p}{m^2}\delta_{[\mu_2}{}^{\nu_1}\dots\delta_{\mu_p}{}^{\nu_{p-1}}\nabla_{\mu_1]}c
\nn & &
 + (-1)^{D-p}\delta_{[\mu_1}{}^{[\nu_1}\dots\delta_{\mu_{2p-D-1}}{}^{\nu_{2p-D-1}}
   u_{\mu_{2p-D}\dots\mu_p]}{}^{\nu_{2p-D}\dots\nu_{p-1}]},
\label{VUbv}
\\
U_{\mu_1\dots\mu_{p-1}}{}^{\nu_1\dots\nu_p} & = &
 \frac{p}{m^2}\delta_{\mu_1}{}^{[\nu_2}\dots\delta_{\mu_{p-1}}{}^{\nu_p}\nabla^{\nu_1]}c
\nn & &
 + \delta_{[\mu_1}{}^{[\nu_1}\dots\delta_{\mu_{2p-D-1}}{}^{\nu_{2p-D-1}}
   u_{\mu_{2p-D}\dots\mu_{p-1}]}{}^{\nu_{2p-D}\dots\nu_p]}.
\label{VUbu}
\eea
\eqref{VUav}, \eqref{VUau} or \eqref{VUbv}, \eqref{VUbu} solve \eqref{O1-2}.
In summary, the solution to \eqref{O1} is
\begin{alignat}{2}
& \bullet~m=0 & \qquad & \text{$\nabla_\rho Y_{(1)}{}^\rho$ is a constant for $D\neq 2p+2$, and $c$ is a constant.}
\notag\\
& \bullet~\text{$m\neq 0$ and $2p\le D$} & \qquad & \text{\eqref{O1-c}, \eqref{VUav}, and \eqref{VUau}}.
\notag\\
& \bullet~\text{$m\neq 0$ and $2p\ge D+1$} & \qquad & \text{\eqref{O1-c}, \eqref{VUbv}, and \eqref{VUbu}}.
\notag
\end{alignat}
Finally we analyze \eqref{O0}. Similarly to the above analysis for \eqref{O1},
we compute the right hand side minus the left hand side of \eqref{O0}.
First we see only the part dependent on $U_{\alpha_1\dots\alpha_{p-1}}{}^{\beta_1\dots\beta_p}$.
The contribution from
$-p\nabla^\rho\nabla_{[\mu_1}s_{|\rho|\mu_2\dots\mu_p]}{}^{\nu_1\dots\nu_p}$
is
\bea
\lefteqn{
-\nabla^\rho\nabla_{[\mu_1}\nabla_{|\rho|}U_{\mu_2\dots\mu_p]}{}^{\nu_1\dots\nu_p}
+(p-1)\nabla^\rho\nabla_{[\mu_1}\nabla_{\mu_2}U_{|\rho|\mu_3\dots\mu_p]}{}^{\nu_1\dots\nu_p}
} \nn & = & 
-\nabla^2\nabla_{[\mu_1}U_{\mu_2\dots\mu_p]}{}^{\nu_1\dots\nu_p}
-\nabla^\rho(\nabla_{[\mu_1}\nabla_{|\rho|}-\nabla_\rho\nabla_{[\mu_1})U_{\mu_2\dots\mu_p]}{}^{\nu_1\dots\nu_p}
\nn & &
+\frac{1}{2}(p-1)\nabla^\rho(\nabla_{[\mu_1}\nabla_{\mu_2}-\nabla_{[\mu_2}\nabla_{\mu_1})
 U_{|\rho|\mu_3\dots\mu_p]}{}^{\nu_1\dots\nu_p}.
\eea
The first term of the second expression of the above cancels the contribution from
 $\nabla^2s_{\mu_1\dots\mu_p}{}^{\nu_1\dots\nu_p}$.
The second and third terms can be simplified by replacing the commutators of covariant derivatives with the Riemann tensors
and using the Bianchi identity.
Then the part dependent on $U_{\alpha_1\dots\alpha_{p-1}}{}^{\beta_1\dots\beta_p}$ is reduced to
\beq
-m^2\nabla_{[\mu_1}{}^\lambda U_{\mu_2\dots\mu_p]}{}^{\nu_1\dots\nu_p}.
\eeq
The part dependent on $V_{\alpha_1\dots\alpha_p}{}^{\beta_1\dots\beta_{p-1}}$ vanishes by the Bianchi identity.
The part dependent on $c$ is
\beq
\delta_{[\mu_1}{}^{\nu_1}\delta_{\mu_2}{}^{\nu_2}\dots\delta_{\mu_p]}{}^{\nu_p}\nabla^2 c
-p\delta_{[\mu_1}{}^{[\nu_1}\delta_{\mu_2}{}^{\nu_2}\dots\delta_{\mu_{p-1}}{}^{\nu_{p-1}}\nabla^{\nu_p]}\nabla_{\mu_p]}c.
\eeq
Finally we see only the part dependent on $Y_{(2n+1)\alpha_1\dots\alpha_n}{}^{\beta_1\dots\beta_n\gamma}$.
The contribution from \\
$-p\nabla^\rho\nabla_{[\mu_1}s_{|\rho|\mu_2\dots\mu_p]}{}^{\nu_1\dots\nu_p}$
needs attention. It is given by
\bea
\lefteqn{
\sum_{n=0}^{p-2}(-1)^{p+n+1}\frac{(n+1)(n+2)}{p-n-1}
 \delta_{[\mu_1}{}^{[\nu_1}\dots\delta_{\mu_n}{}^{\nu_n}\nabla^{\nu_{n+1}}\nabla_{\mu_{n+1}}\nabla_{\mu_{n+2}}
 Y_{(2p-2n-3)\mu_{n+3}\dots\mu_p]}{}^{\nu_{n+2}\dots\nu_p]}
} \nn \lefteqn{
+\sum_{n=0}^{p-1}(-1)^{p+n}\frac{n+1}{p-n}
 \delta_{[\mu_1}{}^{[\nu_1}\dots\delta_{\mu_n}{}^{\nu_n}\nabla^{|\rho|}\nabla_{\mu_{n+1}}\nabla_{|\rho|}
 Y_{(2p-2n-1)\mu_{n+2}\dots\mu_p]}{}^{\nu_{n+1}\dots\nu_p]}
} \nn \lefteqn{
-\sum_{n=0}^{p-2}\frac{(n+1)(p-n-1)}{p-n}
 \delta_{[\mu_1}{}^{[\nu_1}\dots\delta_{\mu_n}{}^{\nu_n}\nabla^{|\rho|}\nabla_{\mu_{n+1}}\nabla_{\mu_{n+2}}
 Y_{(2p-2n-1)\mu_{n+3}\dots\mu_p]\rho}{}^{\nu_{n+1}\dots\nu_p]}
} \nn & = & 
\sum_{n=0}^{p-1}(-1)^{p+n}\frac{n+1}{p-n}
 \delta_{[\mu_1}{}^{[\nu_1}\dots\delta_{\mu_n}{}^{\nu_n}\nabla^2\nabla_{\mu_{n+1}}
 Y_{(2p-2n-1)\mu_{n+2}\dots\mu_p]}{}^{\nu_{n+1}\dots\nu_p]}
\nn & &
+\sum_{n=0}^{p-1}(-1)^{p+n}\frac{n+1}{p-n}
 \delta_{[\mu_1}{}^{[\nu_1}\dots\delta_{\mu_n}{}^{\nu_n}
\nn & &
 \times\nabla^{|\rho|}
 (\nabla_{\mu_{n+1}}\nabla_{|\rho|}-\nabla_{|\rho|}\nabla_{\mu_{n+1}})
 Y_{(2p-2n-1)\mu_{n+2}\dots\mu_p]}{}^{\nu_{n+1}\dots\nu_p]}
\nn & &
+\sum_{n=0}^{p-2}(-1)^{p+n+1}\frac{(n+1)(n+2)}{2(p-n-1)}
 \delta_{[\mu_1}{}^{[\nu_1}\dots\delta_{\mu_n}{}^{\nu_n}
\nn & &
 \times\nabla^{\nu_{n+1}}
 (\nabla_{\mu_{n+1}}\nabla_{\mu_{n+2}}-\nabla_{\mu_{n+2}}\nabla_{\mu_{n+1}})
 Y_{(2p-2n-3)\mu_{n+3}\dots\mu_p]}{}^{\nu_{n+2}\dots\nu_p]}
\nn & &
-\sum_{n=0}^{p-2}\frac{(n+1)(p-n-1)}{2(p-n)}
 \delta_{[\mu_1}{}^{[\nu_1}\dots\delta_{\mu_n}{}^{\nu_n}
\nn & &
 \times\nabla^{|\rho|}
 (\nabla_{\mu_{n+1}}\nabla_{\mu_{n+2}}-\nabla_{\mu_{n+2}}\nabla_{\mu_{n+1}})
 Y_{(2p-2n-1)\mu_{n+3}\dots\mu_p]\rho}{}^{\nu_{n+1}\dots\nu_p]}.
\eea
The first term of the second expression of the above cancels the contribution from
$\nabla^2s_{\mu_1\dots\mu_p}{}^{\nu_1\dots\nu_p}$.
The rest of the above can be simplified by replacing the commutators of covariant derivatives with the Riemann tensors
and using the Bianchi identity.
Then the part dependent on $Y_{(2n+1)\alpha_1\dots\alpha_n}{}^{\beta_1\dots\beta_n\gamma}$ is reduced to
\beq
\frac{2}{D}m^2\nabla_\rho Y_{(1)}{}^\rho\delta_{[\mu_1}{}^{\nu_1}\dots\delta_{\mu_p]}{}^{\nu_p},
\eeq
and \eqref{O0} is reduced to
\bea
0 & = & -m^2\nabla_{[\mu_1} U_{\mu_2\dots\mu_p]}{}^{\nu_1\dots\nu_p}
\nn & &
 +\Big(\nabla^2 c+\frac{2}{D}m^2\nabla_\rho Y_{(1)}{}^\rho\Big)\delta_{[\mu_1}{}^{\nu_1}\dots\delta_{\mu_p]}{}^{\nu_p}
 -p\delta_{[\mu_1}{}^{[\nu_1}\delta_{\mu_2}{}^{\nu_2}\dots\delta_{\mu_{p-1}}{}^{\nu_{p-1}}\nabla^{\nu_p]}\nabla_{\mu_p]}c.
\label{O0-1}
\eea
In the massless case this is satisfied.
In the massive case,  
by contracting $(\mu_1,\dots,\mu_p)$ and $(\nu_1,\dots,\nu_p)$ in \eqref{O0-1}, and using \eqref{O1-aux5},
\beq
m^2\nabla_\rho Y_{(1)}{}^\rho=-\frac{1}{2}(D-2p)\nabla^2 c.
\label{O0-s1}
\eeq
Then we deal with the cases of $2p\le D$ and $2p\ge D+1$ separately.
For $2p\le D$, by contracting $(\mu_2,\dots,\mu_p)$ and $(\nu_2,\dots,\nu_p)$ in \eqref{O0-1},
and using \eqref{VUau}
\beq
\nabla_\mu\nabla_\nu c=\frac{1}{D}g_{\mu\nu}\nabla^2 c.
\label{O0-s2}
\eeq
Then \eqref{O0-s1} and \eqref{O0-s2} solve \eqref{O0-1}. 

For $2p\ge D+1$, by contracting $(\mu_2,\dots,\mu_p)$ and $(\nu_2,\dots,\nu_p)$ in \eqref{O0-1}, and
using \eqref{VUbu}, we obtain \eqref{O0-s2} for $p\le D-2$, and 
for $p=D-1$ (and therefore $D\ge 3$),
\beq
\nabla_\mu\nabla_\nu c-\frac{1}{D}g_{\mu\nu}\nabla^2 c = \frac{(-1)^{D+1}}{D-2}m^2\nabla_\rho u^\rho{}_{\mu\nu},
\eeq
the symmetric part of which also gives \eqref{O0-s2}.
Then for either case, \eqref{O0-1} is reduced to
\beq
0=\delta_{[\mu_1}{}^{[\nu_1}\dots\delta_{\mu_{2p-D-1}}{}^{\nu_{2p-D-1}}
 \nabla_{\mu_{2p-D}}u_{\mu_{2p-D+1}\dots\mu_p]}{}^{\nu_{2p-D}\dots\nu_p]},
\eeq
which is equivalent to
\beq
\nabla_\rho u^\rho{}_{\mu_1\dots\mu_{2D-2p}}=0,
\label{O0-s3}
\eeq
i.e. $u_{\mu_1\dots\mu_{2D-2p+1}}$ is coclosed.

In summary, the solution to \eqref{O0} is
\begin{alignat}{2}
& \bullet~m=0 & \qquad & \text{no new condition}.
\notag\\
& \bullet~\text{$m\neq 0$ and $2p\le D$} & \qquad & \text{\eqref{O0-s1}, and \eqref{O0-s2}}.
\notag\\
& \bullet~\text{$m\neq 0$ and $2p\ge D+1$} & \qquad & \text{\eqref{O0-s1}, \eqref{O0-s2}, and \eqref{O0-s3}}.
\notag
\end{alignat}

Thus we have successfully given a solution to \eqref{Qo3}-\eqref{O0},
only by assuming \eqref{solS}, and that $Y_{(2n+1)\mu_1\dots\mu_n}{}^{\nu_1\dots\nu_n\lambda}$ are 
Killing-Yano forms for $1\le n\le p$ and $Y_{(1)}{}^\lambda$ is a conformal Killing vector.
The solution is summarized as follows:
\bea
S_{\mu_1\dots\mu_p}{}^{\nu_1\dots\nu_p\lambda} & = &
 \delta_{[\mu_1}{}^\lambda U_{\mu_2\dots\mu_p]}{}^{\nu_1\dots\nu_p}
\nn & &
 + \sum_{n=0}^p\delta_{[\mu_1}{}^{[\nu_1}\delta_{\mu_2}{}^{\nu_2}\dots\delta_{\mu_n}{}^{\nu_n}
 Y_{(2p-2n+1)\mu_{n+1}\dots\mu_p]}{}^{\nu_{n+1}\dots\nu_p]\lambda},
\label{psoli}\\
Q_{\mu_1\dots\mu_p}{}^{\nu_1\dots\nu_p\lambda} & = & 
 g^{\lambda[\nu_1}V_{\mu_1\dots\mu_p}{}^{\nu_2\dots\nu_p]}
\nn & &
 + \sum_{n=0}^p\delta_{[\mu_1}{}^{[\nu_1}\delta_{\mu_2}{}^{\nu_2}\dots\delta_{\mu_n}{}^{\nu_n}
 Y_{(2p-2n+1)\mu_{n+1}\dots\mu_p]}{}^{\nu_{n+1}\dots\nu_p]\lambda},
\\
s_{\mu_1\dots\mu_p}{}^{\nu_1\dots\nu_p} & = &
 c\delta_{[\mu_1}{}^{\nu_1}\delta_{\mu_2}{}^{\nu_2}\dots\delta_{\mu_p]}{}^{\nu_p}
 + \nabla_{[\mu_1}U_{\mu_2\dots\mu_p]}{}^{\nu_1\dots\nu_p}
\nn & & 
 + \sum_{n=0}^{p-1}(-1)^{p+n+1}\frac{n+1}{p-n}\delta_{[\mu_1}{}^{[\nu_1}\delta_{\mu_2}{}^{\nu_2}\dots\delta_{\mu_n}{}^{\nu_n}
\nn & & 
 \times\nabla_{\mu_{n+1}}Y_{(2p-2n-1)\mu_{n+2}\dots\mu_p]}{}^{\nu_{n+1}\dots\nu_p]},
\\
q_{\mu_1\dots\mu_p}{}^{\nu_1\dots\nu_p} & = &
 \Big(c+\frac{2}{D}\nabla_\rho Y_{(1)}{}^\rho\Big)\delta_{[\mu_1}{}^{\nu_1}\delta_{\mu_2}{}^{\nu_2}\dots\delta_{\mu_p]}{}^{\nu_p}
\nn & & 
 + \sum_{n=0}^{p-1}(-1)^{p+n+1}\frac{n+1}{p-n}\delta_{[\mu_1}{}^{[\nu_1}\delta_{\mu_2}{}^{\nu_2}\dots\delta_{\mu_n}{}^{\nu_n}
\nn & & 
 \times\nabla_{\mu_{n+1}}Y_{(2p-2n-1)\mu_{n+2}\dots\mu_p]}{}^{\nu_{n+1}\dots\nu_p]},
\eea
where 
\begin{itemize}
\item $m=0$\\
 $\nabla_\rho Y_{(1)}{}^\rho$ is a constant for $D\neq 2p+2$, $c$ is a constant, and
 $U_{\mu_1\dots\mu_{p-1}}{}^{\nu_1\dots\nu_p}$ and $V_{\mu_1\dots\mu_p}{}^{\nu_1\dots\nu_{p-1}}$ are arbitrary tensors
 except that the first $p-1$ and the next $p$ indices of $U_{\mu_1\dots\mu_{p-1}}{}^{\nu_1\dots\nu_p}$, and
 the first $p$ and the next $p-1$ indices of $V_{\mu_1\dots\mu_p}{}^{\nu_1\dots\nu_{p-1}}$ are antisymmetrized.
\item $m\neq 0$\\
 $c$ and $\nabla_\rho Y_{(1)}{}^\rho$ satisfy the followings:
\bea
0 & = & \nabla_\mu\Big(c-\frac{D-2p-2}{2D}\nabla_\rho Y_{(1)}{}^\rho\Big),
\label{sol-mn0-1} \\
m^2\nabla_\rho Y_{(1)}{}^\rho & = & -\frac{1}{2}(D-2p)\nabla^2 c,
\label{sol-mn0-2} \\
\nabla_\mu\nabla_\nu c & = & \frac{1}{D}g_{\mu\nu}\nabla^2 c,
\label{sol-mn0-3} 
\eea
and $U_{\mu_1\dots\mu_{p-1}}{}^{\nu_1\dots\nu_p}$ and $V_{\mu_1\dots\mu_p}{}^{\nu_1\dots\nu_{p-1}}$ are given by
the followings:
\begin{itemize}
\item[$\diamond$] $2p\le D$
\bea
V_{\mu_1\dots\mu_p}{}^{\nu_1\dots\nu_{p-1}} & = &
 \frac{p}{m^2}\delta_{[\mu_2}{}^{\nu_1}\dots\delta_{\mu_p}{}^{\nu_{p-1}}\nabla_{\mu_1]}c,
\\
U_{\mu_1\dots\mu_{p-1}}{}^{\nu_1\dots\nu_p} & = &
 \frac{p}{m^2}\delta_{\mu_1}{}^{[\nu_2}\dots\delta_{\mu_{p-1}}{}^{\nu_p}\nabla^{\nu_1]}c.
\eea
\item[$\diamond$] $2p\ge D+1$
\bea
V_{\mu_1\dots\mu_p}{}^{\nu_1\dots\nu_{p-1}} & = &
 \frac{p}{m^2}\delta_{[\mu_2}{}^{\nu_1}\dots\delta_{\mu_p}{}^{\nu_{p-1}}\nabla_{\mu_1]}c
\nn & &
 + (-1)^{D-p}\delta_{[\mu_1}{}^{[\nu_1}\dots\delta_{\mu_{2p-D-1}}{}^{\nu_{2p-D-1}}
\nn & & 
   \times u_{\mu_{2p-D}\dots\mu_p]}{}^{\nu_{2p-D}\dots\nu_{p-1}]},
\\
U_{\mu_1\dots\mu_{p-1}}{}^{\nu_1\dots\nu_p} & = &
 \frac{p}{m^2}\delta_{\mu_1}{}^{[\nu_2}\dots\delta_{\mu_{p-1}}{}^{\nu_p}\nabla^{\nu_1]}c
\nn & &
 + \delta_{[\mu_1}{}^{[\nu_1}\dots\delta_{\mu_{2p-D-1}}{}^{\nu_{2p-D-1}}
   u_{\mu_{2p-D}\dots\mu_{p-1}]}{}^{\nu_{2p-D}\dots\nu_p]},
\label{psolf}
\eea
where $u_{\mu_1\dots\mu_{2D-2p+1}}$ is an arbitrary coclosed form.
\end{itemize}
\end{itemize}
In addition, we have to note the following fact.
We can consider the Hodge dual $\wt{Y}_{(n)\mu_1\dots\mu_n}$ of $Y_{(D-n)}{}^{\mu_{n+1}\dots\mu_{D}}$:
\beq
\wt{Y}_{(n)\mu_1\dots\mu_n}=\frac{1}{(D-n)!}\epsilon_{\mu_1\dots\mu_D}Y_{(D-n)}{}^{\mu_{n+1}\dots\mu_{D}}.
\eeq
Then if $D-2p+n=0$,
\bea
\lefteqn{
\delta_{[\mu_1}{}^{[\nu_1}\dots\delta_{\mu_n}{}^{\nu_n}Y_{(2p-2n+1)\mu_{n+1}\dots\mu_p]}{}^{\nu_{n+1}\dots\nu_p]\lambda}
} \nn & = & 
 (-1)^{Dp+g+1}\frac{n(D-p)!}{p!}\delta_{[\mu_1}{}^\lambda\epsilon^{\nu_1\dots\nu_p}{}_{\mu_2\dots\mu_{D-p+1}}
 \wt{Y}_{(n-1)\mu_{D-p+2}\dots\mu_p]},
\eea
where $(-1)^g$ is the sign of $g=\det(g_{\mu\nu})$.
Therefore by redefining $U_{\mu_1\dots\mu_{p-1}}{}^{\nu_1\dots\nu_p}$ as
\beq
U_{\mu_1\dots\mu_{p-1}}{}^{\nu_1\dots\nu_p}
\rightarrow U_{\mu_1\dots\mu_{p-1}}{}^{\nu_1\dots\nu_p}
 +  (-1)^{Dp+g}\frac{n(D-p)!}{p!}\epsilon^{\nu_1\dots\nu_p}{}_{[\mu_1\dots\mu_{D-p}}
 \wt{Y}_{(n-1)\mu_{D-p+1}\dots\mu_p]},
\eeq
$Y_{(2p-2n+1)\mu_1\dots\mu_{2p-2n+1}}$ in \eqref{solS} can be absorbed in $U_{\mu_1\dots\mu_{p-1}}{}^{\nu_1\dots\nu_p}$, 
and we can set \\
$Y_{(2p-2n+1)\mu_1\dots\mu_{2p-2n+1}}=0$ as the initial assumption. 
Indeed, in the results in the previous section for $(p,D,n)=(2,3,1), (3,4,2)$ and $(3,5,1)$,
$Y_{(2p-2n+1)\mu_1\dots\mu_{2p-2n+1}}=0$. Then
for $p=1$ and $p=2$ our solution is included in the results in the previous section.
We can also show the followings: If $2p\ge D+1$, then for $0\le n\le 2p-D-1$,
\beq
0=\delta_{[\mu_1}{}^{[\nu_1}\dots\delta_{\mu_n}{}^{\nu_n}Y_{(2p-2n+1)\mu_{n+1}\dots\mu_p]}{}^{\nu_{n+1}\dots\nu_p]\lambda},
\eeq
and for $1\le n\le 2p-D$,
\beq
0=\delta_{[\mu_1}{}^{[\nu_1}\dots\delta_{\mu_{n-1}}{}^{\nu_{n-1}}
 \nabla_{\mu_n}Y_{(2p-2n+1)\mu_{n+1}\dots\mu_p]}{}^{\nu_n\dots\nu_p]}.
\eeq
These can be shown by computing the products of the right hand sides and
$\epsilon^{\mu_1\dots\mu_D}\epsilon_{\nu_1\dots\nu_D}$. 
Therefore if $2p\ge D$, we can set $Y_{(2p-2n+1)\mu_{n+1}\dots\mu_p}{}^{\nu_{n+1}\dots\nu_p\lambda}=0$
for $0\le n\le 2p-D$ as the initial assumption.

For $p=3$ and $D\neq 4,5,6,8$, our assumption from which the above solution is derived is the general solution to
\eqref{Qo3} and \eqref{qo2}. Therefore it gives the general solution to the full conditions \eqref{Qo3}-\eqref{O0}.

From \eqref{sol-mn0-1} and \eqref{sol-mn0-2} we see that $\nabla_\rho Y_{(1)}{}^\rho$ must satisfy
the following ``Klein-Gordon'' equation for $D\neq 2p, 2p+2$:
\beq
0=\nabla^2 (\nabla_\rho Y_{(1)}{}^\rho) + \frac{4D}{(D-2p)(D-2p-2)}m^2\nabla_\rho Y_{(1)}{}^\rho,
\eeq
and for $D=2p$ and $2p+2$, $c$ is a constant and $\nabla_\rho Y_{(1)}{}^\rho=0$.
Furthermore, by acting $\nabla^\nu$ on \eqref{sol-mn0-3}, 
\beq
R_\mu{}^\nu\nabla_\nu c = -\frac{D-1}{D}\nabla_\mu\nabla^2 c.
\eeq
and by applying \eqref{sol-mn0-1} to the left hand side, and \eqref{sol-mn0-2} to the right hand side of this equation,
\beq
(D-2p)(D-2p-2)R_\mu{}^\nu\nabla_\nu(\nabla_\rho Y_{(1)}{}^\rho)
=4m^2(D-1)\nabla_\mu(\nabla_\rho Y_{(1)}{}^\rho).
\eeq
Therefore, if the background metric satisfies the vacuum Einstein equation
\beq
R_{\mu\nu}=\frac{2}{D-2}\Lambda g_{\mu\nu},
\eeq
and 
\beq
m^2\neq \frac{(D-2p)(D-2p-2)}{2(D-1)(D-2)}\Lambda,
\eeq
then $\nabla_\rho Y_{(1)}{}^\rho$ is a constant, and so is $c$. 
After all, in most cases we see that $c$ and $\nabla_\rho Y_{(1)}{}^\rho$ are constants.

\section{Summary and Discussion}

In section 2, we have given the conditions for first order symmetry operators of the second order differential operator
$M_{\mu_1\dots\mu_p}{}^{\nu_1\dots\nu_p}$ which appears in the standard equation of motion for differential $p$-form fields.
Then in section 3, we have given a partial procedure to solve the conditions for arbitrary $p$.
In section 4, we have given the general forms of first order symmetry operators for $p=1$ and $p=2$.
We also have given a partial solution to the conditions for $p=3$.
Then in section 5, based on those solutions we have given a class of first order symmetry operators for arbitrary $p$
and $D(\ge p+1)$. It is summarized in \eqref{psoli}-\eqref{psolf}.
In most cases our symmetry operators commute with $M_{\mu_1\dots\mu_p}{}^{\nu_1\dots\nu_p}$.
We do not have to impose the equation of motion on the background geometry
to show that all of our solutions indeed satisfy the conditions given in section 2.

For $p=1,2$ and $3$,
our solution in section 5 is the general solution for generic $D$. However at some special values of $D$ 
there are more solutions. Then it is natural to conjecture the followings for arbitrary $p$: 
\begin{itemize}
\item For $D\ge 2p+3$ and $D=2p+1$ our solution gives the general solution.
\item For $p+1\le D\le 2p$ our solution is not general.
\item For $D=2p+2$, our solution is not general in the massless case and
 all of $Y_{(2n+1)\mu_1\dots\mu_{2n+1}}$ can be extended to conformal Killing-Yano forms.
\end{itemize}
It is interesting to confirm that these conjectures are correct or need to be corrected.

We have considered the standard linear equations of motion for $p$-forms in purely geometric backgrounds.
However in backgrounds with nontrivial configuration of fields other than the metric,
the equations of motion may have more terms coming from interaction terms in the action.
For such cases our analysis must be modified accordingly.

Most of our symmetry operators consist mainly of odd rank (conformal) Killing-Yano forms. 
Odd dimensional Kerr-NUT-(A)dS spacetimes are examples having such differential forms
(See e.g. \cite{fkk17}.) Indeed the relation between Killing-Yano forms and
the separability of the equations of motion for 1-forms in Myers-Perry-(A)dS or Kerr-NUT-(A)dS spacetimes
is discussed in \cite{l17,fkk18,kfk18,fkks18,hty19} and for $p$-forms in \cite{l19}.
(conformal) Killing-Yano forms also appear in the first order symmetry operators for the equations of motion for
spinor fields (See e.g. \cite{ms79,bc96,bk04,ckk11}), Rarita-Schwinger fields\cite{m18}, and the metric perturbation\cite{m19}.
These facts show the importance of Killing-Yano forms.

\renewcommand{\theequation}{\Alph{section}.\arabic{equation}}
\appendix
\addcontentsline{toc}{section}{Appendix}
\vs{.5cm}
\noindent
{\Large\bf Appendix}
\section{Conformal Killing-Yano forms}
\label{appa}
\setcounter{equation}{0}

A conformal Killing-Yano $n$-form $Y_{\mu_1\mu_2\dots\mu_n}$ is defined by
\beq
\nabla_\nu Y_{\mu_1\mu_2\dots\mu_n} =
 \nabla_{[\nu} Y_{\mu_1\mu_2\dots\mu_n]}
 +\frac{n}{D-n+1}g_{\nu[\mu_1}\nabla_{|\rho|}Y^\rho{}_{\mu_2\dots\mu_n]}.
\label{def:CKY}
\eeq
For $n=1$ this is called a conformal Killing vector. If $Y_{\mu_1\mu_2\dots\mu_n}$ is coclosed i.e.
\beq
\nabla_\rho Y^\rho{}_{\mu_1\dots\mu_{n-1}} = 0,
\eeq
then this is called a Killing-Yano form.
From \eqref{def:CKY},
\bea
\nabla_{\nu_1}\nabla_{\nu_2} Y_{\mu_1\mu_2\dots\mu_n} & = &
 -\frac{1}{2}(n+1)R_{[\nu_2\mu_1|\nu_1}{}^\rho Y_{\rho|\mu_2\dots\mu_n]}
\nn & &
 +\frac{1}{D-n+1}\Big( ng_{\nu_2[\mu_1}\nabla_{|\nu_1}\nabla_{\rho|} Y^\rho{}_{\mu_2\dots\mu_n]}
\nn & &
 - (n+1)g_{\nu_1[\nu_2}\nabla_{\mu_1}\nabla_{|\rho|} Y^\rho{}_{\mu_2\dots\mu_n]}\Big).
\label{ddCKY}
\eea
By antisymmetrizing indices $\nu_1$ and $\nu_2$ in the above, we obtain the following for $n\ge 2$:
\bea
\lefteqn{R_{\nu_1\nu_2[\mu_1}{}^\rho Y_{|\rho|\mu_2\dots\mu_n]}
 + \frac{1}{2}R_{[\mu_1\mu_2|\nu_1}{}^\rho Y_{\rho\nu_2|\mu_3\dots\mu_n]}
 - \frac{1}{2}R_{[\mu_1\mu_2|\nu_2}{}^\rho Y_{\rho\nu_1|\mu_3\dots\mu_n]}
}
\nn & =  &
-\frac{1}{D-n+1}g_{\nu_1[\mu_1}(\nabla_{|\nu_2}\nabla_{\rho|} Y^\rho{}_{\mu_2\dots\mu_n]}
 + \nabla_{\mu_2}\nabla_{|\rho} Y^\rho{}_{\nu_2|\mu_3\dots\mu_n]})
 - (\nu_1\leftrightarrow\nu_2).
\label{CKYint1}
\eea
By contracting $\nu_2$ and $\mu_{n}$, renaming $\mu_{n-1}$ to $\nu_2$, and symmetrizing $\nu_1$ and $\nu_2$
in the above,
we obtain the following for $n\ge 2$:
\bea
0 & = & R_{\nu_1\rho}Y^\rho{}_{\nu_2\mu_1\dots\mu_{n-2}}
 + \frac{1}{2}(n-2)R_{\rho_1\rho_2\nu_1[\mu_1}Y^{\rho_1\rho_2}{}_{|\nu_2|\mu_2\dots\mu_{n-2}]}
\nn & & 
 + \frac{D-n}{D-n+1}\nabla_{\nu_1}\nabla_\rho Y^\rho{}_{\nu_2\mu_1\dots\mu_{n-2}}
 + (\nu_1\leftrightarrow\nu_2).
\label{CKYint2}
\eea

\newcommand{\J}[4]{{\sl #1} {\bf #2} (#3) #4}
\newcommand{\andJ}[3]{{\bf #1} (#2) #3}
\newcommand{\AP}{Ann.\ Phys.\ (N.Y.)}
\newcommand{\MPL}{Mod.\ Phys.\ Lett.}
\newcommand{\NP}{Nucl.\ Phys.}
\newcommand{\PL}{Phys.\ Lett.}
\newcommand{\PR}{Phys.\ Rev.}
\newcommand{\PRL}{Phys.\ Rev.\ Lett.}
\newcommand{\PTP}{Prog.\ Theor.\ Phys.}
\newcommand{\hepth}[1]{{\tt hep-th/#1}}
\newcommand{\arxivhep}[1]{{\tt arXiv.org:#1 [hep-th]}}


\begin{thebibliography}{99}

\bibitem{ms79}
R.\ G.\ McLenaghan, and Ph.\ Spindel,
``{\it Quantum numbers for Dirac spinor fields on a curved space-time}'',
\J{\PR}{D20}{1979}{409}.

\bibitem{bc96}
I.\ M.\ Benn, and P.\ Charlton,
``{\it Dirac symmetry operators from conformal Killing-Yano tensors}'',
\J{Class. Quantum. Grav.}{14}{1997}{1037}, gr-qc/9612011.

\bibitem{bk04}
I.\ M.\ Benn, and J.\ Kress,
``{\it First order Dirac symmetry operators}'',
\J{Class. Quantum. Grav.}{21}{2004}{427}.

\bibitem{ckk11}
M.\ Cariglia, P.\ Krtou\v{s}, and D.\ Kubiz\v{n}\'{a}k,
``{\it Commuting symmetry operators of the Dirac equation, Killing-Yano and Schouten-Nijenhuis brackets}'',
\J{\PR}{D84}{2011}{024004}, arXiv:1102.4501 [hep-th].

\bibitem{m18}
Y.\ Michishita,
``{\it On Quantum Numbers for Rarita-Schwinger Fields}'',
\J{Class. Quantum. Grav.}{36}{2019}{055010}, arXiv:1810.07923 [hep-th].

\bibitem{m19}
Y.\ Michishita,
``{\it First Order Symmetry Operators for the Linearized Field Equation of Metric Perturbations}'',
\J{\PR}{D100}{2019}{124052}, arXiv:1909.12439 [hep-th].

\bibitem{ab16}
S.\ Aksteiner, and T.\ B\"ackdahl,
``{\it Symmetries of linearized gravity from adjoint operators}'',
\J{J. Math. Phys.}{60}{2019}{082501}, arXiv:1609.04584 [gr-qc].

\bibitem{a16}
B.\ Araneda,
``{\it Symmetry operators and decoupled equations for linear fields on black hole spacetimes}'',
\J{Class. Quantum. Grav.}{34}{2017}{035002}, arXiv:1610.00736 [gr-qc].

\bibitem{a17}
B.\ Araneda,
``{\it Generalized wave operators, weighted Killing fields, and perturbations of higher dimensional spacetimes}'',
arXiv:1711.09872 [gr-qc].

\bibitem{l17}
O.\ Lunin,
``{\it Maxwell's Equations in the Myers-Perry Geometry}'',
\J{JHEP}{12}{2017}{138}, arXiv:1708.06766 [hep-th].

\bibitem{fkk18}
V.\ P.\ Frolov, P.\ Krtou\v{s}, and D.\ Kubiz\v{n}\'{a}k,
``{\it Separation of variables in Maxwell equations in Plebanski-Demianski spacetime}'',
\J{\PR}{D97}{2018}{101701}, arXiv:1802.09491 [hep-th].

\bibitem{kfk18}
P.\ Krtou\v{s}, V.\ P.\ Frolov, and D.\ Kubiz\v{n}\'{a}k,
``{\it Separation of Maxwell equations in Kerr-NUT-(A)dS spacetimes}'',
\J{\NP}{B934}{2018}{7}, arXiv:1803.02485 [hep-th].

\bibitem{fkks18}
V.\ P.\ Frolov, P.\ Krtou\v{s}, D.\ Kubiz\v{n}\'{a}k, and J.\ E.\ Santos,
``{\it Massive Vector Fields in Rotating Black-Hole Spacetimes: Separability and Quasinormal Modes}'',
\J{\PRL}{120}{2018}{231103}, arXiv:1804.00030 [hep-th].

\bibitem{fk18}
V.\ P.\ Frolov, and P.\ Krtou\v{s},
``{\it Duality and $\mu$-separability of Maxwell equations in Kerr-NUT-(A)dS spacetime}'',
\J{\PR}{D99}{2019}{044044}, arXiv:1812.08697 [hep-th].

\bibitem{hty19}
T.\ Houri, N.\ Tanahashi, and Y.\ Yasui,
``{\it On symmetry operators for the Maxwell equation on the Kerr-NUT-(A)dS spacetime}'',
\J{Class. Quantum. Grav.}{37}{2020}{015011}, arXiv:1908.10250 [gr-qc].

\bibitem{hty19-2}
T.\ Houri, N.\ Tanahashi, and Y.\ Yasui,
``{\it Hidden symmetry and the separability of the Maxwell equation on the Wahlquist spacetime}'',
\J{Class. Quantum. Grav.}{37}{2020}{075005}, arXiv:1910.13094 [gr-qc]. 

\bibitem{l19}
O.\ Lunin,
``{\it Excitations of the Myers-Perry Black Holes}'',
\J{JHEP}{10}{2019}{030}, arXiv:1907.03820 [hep-th].

\bibitem{fkk17}
V.\ P.\ Frolov, P.\ Krtou\v{s}, and D.\ Kubiz\v{n}\'{a}k,
``{\it Black holes, hidden symmetries, and complete integrability}'',
\J{Living Rev. Relativ.}{}{2017}{20:6}, arXiv:1705.5482 [gr-qc]. 

\end{thebibliography}
\end{document}